\documentclass[AMA,Times1COL]{WileyNJDv5} %STIX1COL,STIX2COL,STIXSMALL

\articletype{Article Type}%

\received{Date Month Year}
\revised{Date Month Year}
\accepted{Date Month Year}
\journal{Journal}
\volume{00}
\copyyear{2023}
\startpage{1}

\usepackage{graphicx}
\usepackage{multirow}
\usepackage{amsmath,amssymb,amsfonts}
\usepackage{amsthm}
\usepackage{mathrsfs}
\usepackage{xcolor}
\usepackage{textcomp}
\usepackage{manyfoot}
\usepackage{booktabs}
\usepackage{algorithm}
\usepackage{algorithmicx}
\usepackage{algpseudocode}
\usepackage{listings}
\usepackage{mathtools}
\usepackage{hyperref}
\usepackage{comment}
\usepackage{adjustbox}
\usepackage{longtable}
\usepackage{lscape}

\makeatletter
\renewcommand{\@cite}[2]{\textsuperscript{#1\if@tempswa , #2\fi}}

\makeatother
\def\trans{^{\rm T}}

\protect
\begin{document}

\title{Least squares-based methods to bias adjustment in scalar-on-function regression model using a functional instrumental variable}

\author[1]{Xiwei Chen}
\author[2]{Ufuk Beyaztas}
\author[1]{Caihong Qin}
\author[1]{Heyang Ji}
\author[3]{Gilson Honvoh}
\author[1]{Roger S. Zoh}
\author[4]{Lan Xue}
\author[1]{Carmen D. Tekwe}

\authormark{CHEN \textsc{et al.}}
\titlemark{Least squares-based methods to bias adjustment in scalar-on-function regression model using a functional instrumental variable}

\address[1]{\orgdiv{Department of Epidemiology and Biostatistics}, \orgname{Indiana University, School of Public Health}, \orgaddress{\state{Indiana}, \country{US}}}
\address[2]{\orgdiv{Department of Statistics}, \orgname{Marmara University}, \orgaddress{\country{Turkey}}}
\address[3]{\orgdiv{Department of Pediatrics}, \orgname{Cincinnati Children's Hospital Medical Center}, \orgaddress{\state{Ohio}, \country{US}}}
\address[4]{\orgdiv{Department of Statistics}, \orgname{Oregon State University}, \orgaddress{\state{Oregon}, \country{US}}}

\corres{Corresponding author Carmen D. Tekwe, 1025 E. 7th Street, Bloomington, Indiana, US. \email{ctekwe@iu.edu}}

\presentaddress{1025 E. 7th Street, Bloomington, Indiana, US.}

\abstract[Abstract]{Instrumental variables (IVs) are widely used to adjust for measurement error (ME) bias when assessing associations of health outcomes with ME-prone independent variables. IV approaches addressing ME in longitudinal models are well established, but few methods exist for functional regression. We develop two methods to adjust for ME bias in scalar-on-function linear models. We regress a scalar outcome on an ME-prone functional variable using a functional IV for model identification and propose two least squares–based methods to adjust for ME bias. Our methods alleviate potential computational challenges encountered when applying classical regression calibration methods for bias adjustment in high-dimensional settings and adjust for potential serial correlations across time. Simulations demonstrate faster run times, lower bias, and lower AIMSE for the proposed methods when compared to existing approaches. The proposed methods were applied to investigate the association between body mass index and wearable device-based physical activity intensity among community-dwelling adults living in the United States.}


\jnlcitation{\cname{%
\author{Chen X.},
\author{Beyaztas U.},
\author{Qin C.},
\author{Ji H.},
\author{Honvoh G.},
\author{Zoh R.S.},
\author{Xue L.}, and
\author{Tekwe C.D.}}.
\ctitle{Least squares-based method to bias adjustment in scalar-on-function regression model using a functional instrumental variables.} \cjournal{\it XXXXXXXXXXXXX.} \cvol{0000;00(00):0--00}.}

\maketitle

\section{Introduction}\label{sec:introduction}
Measurement error occurs in observational and epidemiological studies when the interest lies in monitoring or measuring exposures that are difficult to measure directly either due to prohibitive costs or the exposure being latent \cite{carroll2006measurement, fuller2009measurement}. When true exposures are latent or difficult to measure directly, proxy or surrogate measures that represent their unbiased measures are used instead. For example, when assessing how usual dietary intake influences health outcomes in nutrition epidemiological studies, self-reported measures of dietary intake based on responses to food frequency questionnaires are often used \cite{kristal2005time,cade2002development,spiegelman1997regression, carroll2006measurement,cade2004food}. Self-reported measures of physical activity patterns based on responses to the Global Physical Activity Questionnaire have also been used as proxy measures for physical activity patterns \cite{welk2023equating,armstrong2006development,troiano2020can}. Although these self-reported proxy measures provide an estimate of the exposure of interest, they have been well documented to be prone to measurement errors due to recall bias \cite{carroll2006measurement, spiegelman1997regression,tooze2013measurement,tekwe2014multiple}. For example, under-reporting of dietary intake in 24-hour recall and food frequency questionnaires can range from 10\% to 50\%, highlighting the potential effects of recall bias and day-to-day variation on accurate dietary intake assessments \cite{macdiarmid1998assessing,park2007underreporting,subar2003using}. 

To address some of these limitations in self-reported measures of exposures such as physical activity behavior or usual dietary intake, more objective measures are used. Wearable devices are increasingly used to monitor physical activity and collect data at frequent intervals such as in epochs of 60 seconds over multiple days, resulting in high-dimensional curves of functional data \cite{ramsay2005functional,ferraty2006nonparametric,yao2005functional}. Although the use of wearable devices provides improved data accuracy compared to self-reported measures, their accuracy has also been questioned \cite{tekwe2018functional,warolin2012effect,matthews2018influence,jacobi2007physical,bassett2012device,rothney2008validity}. Some of the factors that induce potential measurement error biases with wearable devices for physical activity include sensor accuracy, device placement, activity algorithm, and user behavior \cite{fuller2020reliability,xie2018evaluating,chowdhury2017assessment,williams2023wearable,arvidsson2019measurement}. 

There is a vast amount of literature and work devoted to addressing measurement error for scalar-valued exposures in regression models \cite{fuller2009measurement,carroll2006measurement}, however, less work has been devoted to addressing measurement error when the exposure appears as functions. Some of the factors that make the correction of the measurement error in functional exposures more complex compared to the measurement error in scalar-valued exposures include the high-dimensional nature of functional data and the complex heteroskedastic measurement error structures associated with the frequency of data collection \cite{tekwe2019instrumental}. Some of the commonly used approaches for addressing measurement error biases include the assumption of a known covariance matrix for measurement error, repeated measurements on the proxy measures, or the availability of validation data \cite{hu2008instrumental, carroll1994measurement,spiegelman1997regression,bollen2012instrumental,fuller2009measurement}. In this manuscript, we propose and demonstrate the use of function-valued instrumental variables to adjust for biases due to measurement error in functional linear regression. 

Instrumental variable-based methodology has been widely applied in various settings, including social sciences, econometrics, and biostatistics \cite{newhouse1998econometrics, angrist1991instrumental, widding2022gentle, chernozhukov2007instrumental, bollen2012instrumental,zoh2024bayesian,tekwe2019instrumental}. Certain conditions are required for a variable to be considered an instrument. The first condition is relevance. The relevance condition requires that an instrument exhibits a significant correlation with the endogenous explanatory variable or exposure \cite{bollen2012instrumental}. This relationship is needed for the instrument to serve as a viable proxy for the exposure or covariate prone to measurement error, facilitating the isolation of the true effect of the covariate on the outcome \cite{pokropek2016introduction, pearl2000models}. Without this correlation, the instrument would fail to provide the necessary leverage to unravel the impact of the endogenous variable from the confounding effects of measurement error. The second condition is the exogeneity condition \cite{bollen2012instrumental}. The exogeneity condition requires that the instrument not be correlated with the measurement error that affects the explanatory variable. This condition is essential to ensure that the instrument does not introduce additional bias into the analysis. The instrumental variable's exogeneity guarantees that any correlation between the instrument and the outcome variable operates solely through the endogenous explanatory variable, thus preserving the integrity of the instrumental variable estimation \cite{carroll2006measurement,bollen2012instrumental,greenland2000introduction,pokropek2016introduction, angrist2001instrumental}.

Tekwe et al. (2019) proposed the use of a functional instrumental variable to correct for measurement error in functional linear regression \cite{tekwe2019instrumental}. Under this approach, a generalized method of moment-based approach was developed to estimate and account for the complex heteroscedastic structures associated with the measurement error and improved on the biases associated with naive approaches that do not adjust for measurement errors \cite{tekwe2019instrumental}. The authors also demonstrated that formally addressing the biases due to measurement error in functional linear regression models outperformed smoothing-based approaches that are often used for measurement error correction in traditional functional regression models \cite{tekwe2019instrumental}. In another manuscript, Tekwe et al. (2022) developed an approach to address measurement error in functional quantile regression models when assessing the association between an error-prone functional exposure and the quantile functions of a continuous response. Estimation was based on the simulation extrapolation method (SIMEX) \cite{tekwe2022estimation}. More recently, Zoh et al. developed a fully Bayesian approach to the estimation of the functional linear regression model using an instrumental variable \cite{zoh2024bayesian}. The authors compared the fully Bayesian semiparametric methods with frequentist-based approaches. 

In this manuscript, we propose two regression-calibration-based methods that employ ordinary least squares with an instrumental variable to correct for estimation biases resulting from measurement error. In the first method, we modify the univariate pointwise approach proposed by Cui et al. (2022) to adjust for the measurement error bias in estimation by using a functional instrumental variable \cite{cui2022fast}. However, the first method overlooks serial correlations within subjects at multiple observation time points. To address this limitation, we developed a new method that utilizes the complete data set across all observation periods. In the second method, we first implement dimensionality reduction in the functional data and develop a multivariate framework to account for measurement error. Additionally, we adapt our previously described SIMEX method for functional quantile regression models to functional linear regression for comparison. We applied our proposed methods to assess the association between device-based measures of weekday physical activity and body mass index (BMI) among community-dwelling adults living in the United States using the data from the National Health and Nutrition Examination Survey (NHANES). We consider device-based measures of weekday physical intensity as measurement error-prone proxies of true weekday physical activity intensity, while the device-based measures of weekend physical intensity are used as the functional instrumental variable. We also established some finite sample properties of our proposed methods through simulations.

The remainder of this manuscript is organized as follows. In Section~\ref{sec:methods}, we provide an overview of the functional linear regression model and introduce our proposed measurement error adjustment approaches. Section~\ref{sec:simulations} presents the results of various Monte Carlo experiments carried out under different data generation scenarios to evaluate the finite-sample performance of the proposed methods. In Section~\ref{sec:application}, we apply the methods to real-world data on adult health and activity levels, showcasing their practical utility. Finally, Section~\ref{sec:discussion} concludes the paper with a discussion of possible extensions of the methodology.

\section{Methods}\label{sec:methods}

\subsection{Functional Linear Regression Model}\label{subsec:model}
Let $\{Y_{i},X_{i}(t),Z_{i}\}_{i=1}^{n}$ be a triplet consisting of a continuous scalar-valued outcome $Y_{i}$; a random functional-valued covariate $X_{i}(t)$; and a $p \times 1$ vector of scalar-valued error-free covariates $Z_{i}$ for the $i$-{th} subject, respectively. Here, $X_i(t)$, $t \in \mathcal{I}$, represents a functional covariate exhibiting complex temporal behavior, where $\mathcal{I}$ is a compact interval defining the domain of interest. This domain $\mathcal{I}$ provides the framework within which processes $X_i(t)$s  are studied for $i =1, \ldots, n$. In this study, without loss of generality, we assume that $\mathcal{I} = [0,1]$. In addition, we postulate that $X_{i}(t)$ is a latent unobservable covariate for subject $i$ at time $t$. Let $W_{i}(t)$ denote an imprecisely observed proxy measure for $X_{i}(t)$ prone to measurement error, denoted by $U_{i}(t)$. Then, the functional linear regression model with measurement error for subject $i$ at time $t$ is given as follows:
\begin{eqnarray}
{Y_{i}} &=& \beta_{0} + \int_{0}^{1} \beta_{1}(t)X_{i}(t)dt + \gamma Z_{i}\trans + \varepsilon_{i},  \label{eq1}  \\
W_{i}(t) &=&  X_{i}(t) + U_{i}(t),  \label{eq2}
\end{eqnarray}
where $\beta_{0}$ is a fixed intercept, $\beta_{1}(t)$ is a functional coefficient that relates the latent exposure $X_{i}(t)$, to the response $Y_{i}$, $\gamma$ is a $p \times 1$ vector of coefficients for the error-free covariates, and $\varepsilon_{i}$ is the model error with $\varepsilon \sim N(0, \sigma^2)$. Let $\Sigma_{UU}$ be the covariance matrix of $U_{i}(t)$. In~\eqref{eq1} and~\eqref{eq2}, we impose the assumptions that $E(U_{i}|X_{i}) = 0$ and $\mathrm{Cov}(X_{i}, U_{i}) = 0$.  Finally, we also assume that the error terms, $\varepsilon_{i}$ and $U_{i}(t)$ are independent.

% Basis expansion
To reduce the dimension of the functional term in~\eqref{eq1}, a polynomial spline smoothing method is applied. Following the polynomial spline basis expansion, the functional coefficient in~\eqref{eq1} is approximated as $\beta_{1}(t)\approx \sum_{k=1}^{K}\omega_{k}b_{k}(t)$, where $b_{k}$ are a set of spline basis functions and $\omega_{k}$ are the associated spline coefficients, and $K$ represents the number of spline basis. We select $K$ by employing the Bayesian Information Criterion (BIC) to capture the intrinsic characteristics of the original data \cite{kato2012estimation}. Specifically, we fit the model for a range of $K$ values and select the optimal $K$ that minimizes the BIC, ensuring a balance between model complexity and the goodness of fit. With the optimum $K$, the models in~\eqref{eq1} and~\eqref{eq2} are expressed as follows:
\begin{eqnarray}
{Y_{i}} &\approx& \beta_{0} + \sum_{k=1}^{K}\omega_{k}X_{ik} + \gamma Z_{i}\trans + \varepsilon_{i},  \label{eq3}  \\
W_{ik} &=& X_{ik} + U_{ik}, \quad k = 1,\ldots,K,  \label{eq4}
\end{eqnarray}
where $X_{ik} = \int_{0}^{1}X_{i}(t)b_{k}(t)dt$, $W_{ik} = \int_{0}^{1}W_{i}(t)b_{k}(t)dt$, and $U_{ik} = \int_{0}^{1}U_{i}(t)b_{k}(t)dt$.

In our study, the BIC is defined based on the observed measurements \{$W_{i}(t)$\} as 
\begin{equation*}
BIC(K) = log \left[\frac{1}{n} \sum_{i=1}^{n} (Y_{i} - {\beta}_{0} - Z_{i}\trans\hat{\gamma} - \sum_{k=1}^{K} W_{ik}\hat{\omega}_{k})^2 \right] + \frac{(K+p)log(n)}{n}.
\end{equation*}
We acknowledge that the BIC used in this study is not a perfect BIC as $X_i(t)$ is not practically available.

Given that $X_i(t)$ is an unobservable latent exposure, replacing it with $W_{i}(t)$, its observed measure, in~\eqref{eq1} to assess the association between $X_i(t)$ and $Y_i$ would lead to biased estimates for $\beta_{1}(t)$ due to the presence of the measurement errors, $U_i(t)$.\cite{carroll2006measurement, fuller2009measurement}

\subsection{Measurement Error Adjustment Approaches}\label{subsec:me_approach}
To address estimation bias due to measurement error, some additional identifying information is needed to estimate the covariance function associated with the measurement error, $\Sigma_{UU}$. Sources of additional identifying information include repeated measures on $W_{i}(t)$, internal or external validation data, or instrumental variables \cite{carroll2006measurement, fuller2009measurement}. Once the additional data has been identified, various estimation procedures may be implemented to estimate $\beta_{1}(t)$ in~\eqref{eq1}. Some of these estimation approaches include regression calibration, SIMEX, method-of-moments, and corrected scores \cite{spiegelman1997regression, hardin2003regression, spiegelman2011regression, stefanski1995simulation, carroll1997asymptotic, luan2023generalized, polzehl2004symmetrized, cook1994simulation, nakamura1990corrected, cowan1998method}. 

In our current work, we propose two different two-stage least squares (2SLS) methods that use a functional instrumental variable as additional identifying information with regression calibration for estimation; as well as a SIMEX method with an instrumental variable. The 2SLS method is widely used to address endogeneity and measurement error biases in economics with scalar covariates \cite{angrist1991does, card1993using, oreopoulos2006intergenerational, wooldridge2016introductory}, and its extension to functional data is novel. In 2SLS-based approaches, a functional instrumental variable, $M_{i}(t)$, together with the functional proxy measure, $W_{i}(t)$, is used to obtain estimates of the true exposure, $X_{i}(t)$. The true exposure is then replaced with its estimated values for use in the outcome regression model in~\eqref{eq1}.

For $i =1,\ldots,n$, let $M_{i}(t)$ represent a function-valued instrumental variable for $X_{i}(t)$. We introduce the model for $M_i(t)$ as follows:
\begin{eqnarray}
M_{i}(t) &=& \delta(t)X_{i}(t) + \eta_{i}(t),  \label{eq5}
\end{eqnarray}
where $\delta(t)$ is an unknown functional coefficient and $\eta_{i}(t)$ is the model error. A valid instrumental variable relies on the following key assumptions:
\begin{enumerate}
\item $M_{i}$ is linearly correlated with $X_{i}$ with $\delta(t) \neq 0$.

\item $M_{i}$ is uncorrelated with the measurement error with $\mathrm{Cov}\left(M_{i}(t), U_{i}(s)\right) = 0$ for any $t,s\in \mathcal{I}$.
\end{enumerate}
These two assumptions ensure that the instrumental variable, $M_{i}$, has the necessary correlation with the true functional covariate, $X_i(t)$, but remains independent of measurement errors, allowing us to establish unbiasedness in the estimation.

\subsubsection{Pointwise two-stage least squares approach (PW-2SLS)}\label{subsubsec:pw2sls}
In Cui et al. (2022), the univariate pointwise method was proposed as a fast and scalable approach to fit multi-level functional mixed-effects models while reducing the computational burden associated with the regression analysis of longitudinal functional data \cite{cui2022fast}. This approach analyzes the data by applying a linear mixed effects model at each wear time across multiple days. Instead of repeated measures of the functional data, we aim to use a functional instrumental variable for model identification. Our univariate two-stage least squares method analyzes the data with one wear time at a time across all wear times with the following steps.

\begin{enumerate}
\item A univariate pointwise linear regression model is performed at each time point $t$, associated with the functional terms. \\
Thus, for each time $t$, we regress $W_{i}(t)$ on $M_{i}(t)$ with
\begin{eqnarray}
W_{i}(t) = \alpha_{0}(t) + \alpha_{1}(t)M_{i}(t) + \zeta_{i}(t),  \label{eq6}
\end{eqnarray}
The regression coefficients $\alpha_0(t)$ and $\alpha_1(t)$ for each time can be estimated by ordinary least squares. That is, for any fixed $t$,
\begin{equation*}
\left(\widehat{\alpha}_{0}(t),\widehat{\alpha}_{1}(t)\right) = \arg \min_{\alpha_{0}(t),\alpha_{1}(t)} \sum_{i=1}^{n} \{W_{i}(t) - \alpha_{0}(t) - \alpha_{1}(t) M_{i}(t)\}^{2}.
\end{equation*}

\item The fitted values are obtained from~\eqref{eq6} across all time points as $\hat W_{i}(t)=\widehat{\alpha}_{0}(t)+\widehat{\alpha}_{1}(t)M_{i}(t)$, and use them as plugged values for the functional covariate in the main regression model.

\item The dimension of $\hat W_{i}(t)$ is reduced by using the polynomial basis expansion method described in Section \ref{subsec:model} to obtain $\hat W_{ik}$. 

\item Lastly, replace $X_{ik}$ with $\hat W_{ik}$ in~\eqref{eq3} to estimate $\hat \omega_{k}$ and the functional coefficient in~\eqref{eq1} is estimated as $\hat \beta_{1}(t) = \sum_{k=1}^{K}\hat \omega_{k}b_{k}(t)$.
\end{enumerate}

\subsubsection{Multivariate two-stage least squares approach (MULTI-2SLS)}\label{subsubsec:multi2sls}
The previously described pointwise two-stage least squares estimation method conducts the analysis separately for each time point. As a result, the PW-2SLS approach cannot adequately account for the correlation between observations at each time point $t$. To address this limitation, a novel method was developed using a multivariate framework by considering all time points of the instrumental variable and the proxy measure simultaneously in the first stage. The multivariate two-stage least squares approach approach is described below.

\begin{enumerate}
\item First, reduce the dimension of the functional variables $W_{i}(t)$ and $M_{i}(t)$, using the polynomial spline basis expansion method described in Section \ref{subsec:model} and get $W_{ik}$ and $M_{ik}$ as 
\begin{equation*}
W_{i}(t) \approx \sum_{k=1}^K W_{ik}b_{k}(t), \quad M_{i}(t) \approx \sum_{k=1}^K M_{ik}b_{k}(t).
\end{equation*}

\item A multivariate linear regression model is performed for each basis of $W_{ik}$. That is, for each $k$, we regress $W_{ik}$ on $M_{i1},\ldots, M_{iK}$ and consider
\begin{equation}
W_{ik} = \alpha_{k0} + \sum_{j=1}^{K}\alpha_{kj}M_{ij} + \zeta_{ik},  \label{eq7}
\end{equation}
Let ${\bf \alpha}_{k}=\left(\alpha_{k0},\alpha_{k1},\ldots,\alpha_{kK}\right)^{T}$ be the regression coefficients, which can be estimated by
\begin{equation*}
\widehat{\alpha}_{k} = \arg \min_{\alpha_k} \sum_{i=1}^{n} \{W_{ik} -\alpha_{k0} - \sum_{j=1}^{K} \alpha_{kj}M_{ij}\}^{2}
\end{equation*}

\item Obtain the fitted values from~\eqref{eq7} across all basis as $\hat W_{ik}=\hat{\alpha}_{k0} + \sum_{j=1}^{K} \hat{\alpha}_{kj}M_{ij}$.

\item Replace $X_{ik}$ with $\hat W_{ik}$ in~\eqref{eq3} to estimate $\hat \omega_{k}$ and the functional coefficient in~\eqref{eq1} is estimated as $\hat \beta_{1}(t) = \sum_{k=1}^{K}\hat \omega_{k}b_{k}(t)$. 
\end{enumerate}

\subsubsection{Simulation Extrapolation (SIMEX) Approach}\label{subsubsec:simex}
We also applied the SIMEX method previously described by Tekwe et al. \cite{tekwe2022estimation} for functional quantile regression models to this functional linear regression setting and compared its performance with the regression calibration-based approaches. The SIMEX method involves a simulation step followed by an extrapolation step. The detailed steps involved in implementing the SIMEX method are summarized as follows \cite{stefanski1995simulation,tekwe2022estimation}:

\begin{enumerate}
\item Estimate $\delta(t)$ using the following data-driven ratio estimator:
\begin{equation*}
\widehat\delta(t) = \frac{\frac{1}{n} \sum_{i=1}^{n} M_{i}(t)}{\frac{1}{n} \sum_{i=1}^{n} W_{i}(t)}.    
\end{equation*}
And~\eqref{eq5} can be written as follows:
\begin{equation*}
M_{i}^\ast(t) = X_{i}(t) + \eta_{i}^\ast(t),
\end{equation*}
where $M_{i}^\ast(t) = M_{i}(t)/\widehat\delta(t)$ and $\eta_{i}^\ast(t) = \eta_{i}(t)/\widehat\delta(t)$.

\item Apply the basis expansion to $M_{i}^\ast(t)$ and $W_{i}(t)$ and estimate the covariance matrix of the measurement error {$U_{i}(t)$} as follows:
\begin{equation*}
\widehat{\Sigma}_{UU} = \widehat{\Sigma} - \widehat{\Sigma}_{M*W},
\end{equation*}
where $\Sigma_{WW}$ is the covariance matrix of $W_{ik}$ and $\Sigma_{M*W}$ is the covariance matrix between $M_{ik}^\ast$ and $W_{ik}$.

\item Choose a set of small numbers greater than 0 that increase monotonically to represent $\lambda_{l}$, $l \in \{1, \ldots, L$\}. The $\lambda_{l}$'s are used as the scale for the simulated errors added to the data. For example, $\lambda = \{0.0001, 0.0501, 0.1001, \ldots, 2.0001$\}.

\item Simulate additional measurement errors and add them to the proxy measure:
\begin{equation*}
W(\lambda_{l}) = W + \sqrt{\lambda_{l}}U \quad l \in \{1, \ldots, L\},
\end{equation*}
where $U \sim MVN(0, \widehat{\Sigma}_{UU})$.

\item Estimate the beta coefficients ($\widehat{\omega}_{k}(\lambda_{l})$) of the linear regression model of Y using the simulated data $W(\lambda_{l})$.

\item Iterate between the simulation and estimation steps (Steps 3 and 4) multiple times and compute the average values of the beta coefficients ($\widehat{\omega}_{k}(\lambda_{l})$) over all iterations.
\begin{equation}
\widehat{\omega}_{k}(\lambda) = \frac{1}{L} \sum_{l=1}^{L} {\omega}_{k}(\lambda_{l})
\end{equation}

\item Extrapolate the results from Step 5 and estimate the functional coefficient as follows:
\begin{equation*}
\widehat{\beta}_1(t) = \sum_{k=1}^{K} \widehat{\omega}_{k}(\lambda=-1)b_{k}(t),
\end{equation*}
where $\lambda=-1$ corresponds to the scenario of no measurement error. $\lambda=0$ represents using $W$ to estimate $\beta_{1}(t)$ directly.
\end{enumerate}

\section{Simulations}\label{sec:simulations}
We conducted five simulation studies to evaluate the performance of our proposed methods. In each set of simulations, we generated 500 independent data sets. The true function-valued covariate was simulated as $X(t) = 1/[1+\exp\{8 (t-0.5)\}]+1 + \varepsilon_{X}(t)$, where the error term $\varepsilon_{X}(t)$ follows a multivariate Gaussian distribution with $MVN(0, \Sigma_{X})$. The function-valued covariate prone to measurement error was simulated as $W(t) = X(t) + U(t)$. The function-valued instrumental variable was simulated as $M(t) = \delta(t)X(t) + \eta(t)$, where $\delta(t) = c \sin(2 \pi t)+1$ with $c$ being a constant, and the error term $\eta(t) \sim MVN(0, \Sigma_{M})$. Different variance-covariance structures for functional terms were considered. We also included two scalar-valued error-free covariates in the model, a continuous one, $Z_{c} \sim N(0, 0.5^{2})$, and a binary one, $Z_{b} \sim B(n, 0.6)$. Finally, the scalar-valued outcome $Y$ was generated as 
\begin{equation*}
Y = \int_{0}^{1} \sin(2 \pi t) X(t)dt + 2 Z_{c} + 0.6 Z_{b} + \varepsilon,
\end{equation*}
where $\varepsilon \sim N(0, 0.1^2)$.
%We considered time points $t \in \{1, \ldots, 100 \}$ for the functionalcovariate. 

Prior to implementing the measurement error adjustment approaches, we selected the optimal number of knots for basis expansion using the BIC method. We initially set the lower and upper bounds for $K$ to $5$ and $9$, respectively, and the optimal number of knots is the one that minimizes the BIC value. Note that the upper and lower bounds of $K$ were chosen arbitrarily.

For a sequence of equally spaced grid points on $[0,1]$, $\left\{t_{l}\right\}_{l}^{n_{grid}}$, the finite-sample performance of the methods was assessed using three measures: average squared bias ($\text{ABias}^{2}$), average sample variance ($\text{AVar}$), and the average integrated mean square error (AIMSE) as
\begin{align*}
\text{ABias}^2\{\widehat{\beta}_{1}(t)\} &= \frac{1}{n_{grid}} \sum_{l=1}^{n_{grid}} \{\bar{\beta}_{1}(t_{l}) -{{{\beta_{1}(t_{l})}}}\}^{2}, \\
\text{AVar}\{\widehat{\beta}_{1}(t)\} &= \frac{1}{500}\sum_{r=1}^{500}\frac{1}{n_{grid}}\sum_{l=1}^{n_{grid}}\left\{\widehat{\beta}_{1}^{(r)}(t_{l})-\bar{{\beta}}_{1}(t_{l})\right\}^{2}, \\
\text{AIMSE}\{\widehat{\beta}_{1}(t)\} &= ABias^2\{\widehat{\beta}_{1}(t)\} + AVar\{\widehat{\beta}_{1}(t)\},
\end{align*}
in which $\bar{\beta}_{1}(t) = \frac{1}{500} \sum_{r=1}^{500} \beta_{1}^{(r)}(t)$.

Furthermore, we obtained the mean squared percent error (MSPEE) between the true and estimated values of $\beta_{1}(t)$. 
\begin{equation*}
\text{MSPEE} = 100 \times \sqrt{\frac{\int_0^1 (\beta_{1}(t) - \hat{\beta}_{1}(t))^2 \, dt}{\int_0^1 \beta_{1}(t)^2 \, dt}}
\end{equation*}

We compared the performance of five estimators in estimating the true functional coefficient $\beta_{1}(t)$. These estimators include $(1)$ Oracle estimator (Oracle), obtained by using the true predictor $X(t)$ for estimation, which will be used as our benchmark; $(2)$ MULTI-2SLS; $(3)$ PW-2SLS; $(4)$ SIMEX; and $(5)$ Naive, which uses the proxy measure $W(t)$ to estimate the effects of $X(t)$ on the response, ignoring the measurement error.

\subsection{Simulation Study 1}
In the first simulation study, we examined the effect of sample sizes on estimating the true functional coefficient $\beta_{1}(t)$. The structures of the covariance function for the functional-valued covariates \{i.e., $X(t)$, $W(t)$, and $M(t)$\} were restricted as autoregressive of order one \{$AR(1)$\} with correlations of $\rho_{X} = \rho_{W} = \rho_{M} = 0.5$. The error term of $X(t)$ was generated as $\varepsilon_{X}(t) \sim MVN(0, \Sigma_{X})$ with $\Sigma_{X} = 1.5$. The measurement error was simulated as $U(t) \sim MVN(0, \Sigma_{U})$ with $\Sigma_{U} = 1$. The instrumental variable $\{M(t)\}$ was simulated with $M(t) = \delta(t)X(t) + \eta(t)$, where $\delta(t) = 0.5 \sin(2 \pi t)+1$ and $\eta(t) \sim MVN(0, \Sigma_{M})$ with $\Sigma_{M} = 1$. We considered sample sizes of $n=(100, 500, 1000, 5000)$.

Table~\ref{table:samplesize} shows the performance comparison of the five estimation methods at varying sample sizes. Aside from the Oracle estimator, which consistently exhibited the lowest bias, the MULTI-2SLS and PW-2SLS estimators followed with the next lowest biases, showing significant improvement as the sample size increased. The biases of the SIMEX estimators were consistently higher than those of MULTI-2SLS and PW-2SLS estimators across all sample sizes. In terms of variance, all estimators began with high values that steadily decreased to small, comparable levels. The MULTI-2SLS, PW-2SLS, and SIMEX estimators demonstrated similar variances. For AIMSE, the Oracle estimator consistently outperformed all others. MULTI-2SLS and PW-2SLS estimators ranked next, with AIMSE decreasing as the sample sizes increased. SIMEX performed moderately, while the Naive estimator consistently had the highest AIMSE.

In general, as the sample size increased, the performance of all estimators improved. Oracle estimators consistently outperformed the others across all metrics. The least squares-based methods showed strong results, particularly with larger samples. The SIMEX approach also performed competitively. In contrast, the Naive method had the poorest performance, highlighting its unsuitability for accurate estimation.

Furthermore, the estimation of coefficients for scalar error-free covariates ($\gamma$) showed a similar pattern, all estimators improved in performance as the sample size increased, with biases nearing zero and variances/AIMSE decreasing consistently. Among the estimators, the Oracle estimator generally achieved the best performance, whereas the MULTI-2SLS and PW-2SLS estimators exhibited lower biases, and the SIMEX and Naive estimators showed lower variances. These findings underscore that increasing the sample size enhances the accuracy of estimators and reduces error variability across all evaluated metrics.

\begin{table}[ht]
\centering\tiny\setlength\tabcolsep{3.5pt}
\caption{The influence of sample sizes on the performance of five estimators (Oracle, MULTI-2SLS, PW-2SLS, SIMEX, and Naive) in estimating $\beta_{1}(t)$. The functional variables were simulated under AR(1) covariance structures with $\rho_{X} = \rho_{W} = \rho_{M} = 0.5$, $\varepsilon_{X}(t) \sim MVN(0, \Sigma_{X})$ with $\Sigma_{X} = 1.5$, $U(t) \sim MVN(0, \Sigma_{U})$ with $\Sigma_{U} = 1$, $\eta(t) \sim MVN(0, \Sigma_{M})$ with $\Sigma_{M} = 1$, and $\delta(t) = 0.5 \sin(2 \pi t)+1$.}
\label{table:samplesize}
\centering
\begin{tabular}{r|ccccc}
\hline
&\multicolumn{5}{c}{$\text{ABias}^{2}$ (MSPEE, \%)}\\
\hline
n&Oracle&MULTI-2SLS&PW-2SLS&SIMEX&Naive\\
\hline
100 & 0.0030 (7.4) & 0.0027 (7.0) & 0.0028 (7.2) & 0.0083 (12.8) & 0.0485 (31.3)\\
500 & 0.0019 (5.8) & 0.0016 (5.5) & 0.0018 (5.8) & 0.0062 (11.0) & 0.0484 (31.2)\\
1000 & 0.0004 (2.7) & 0.0003 (2.4) & 0.0004 (2.5) & 0.0048 ( 9.8) & 0.0476 (31.0)\\
5000 & 0.0001 (1.1) & 0.0001 (1.1) & 0.0001 (1.1) & 0.0040 ( 9.0) & 0.0468 (30.7)\\
\hline
&\multicolumn{5}{c}{$\text{AVar}$}\\
\hline
n&Oracle&MULTI-2SLS&PW-2SLS&SIMEX&Naive\\
\hline
100 & 0.0108 & 0.0438 & 0.0342 & 0.0311 & 0.0151\\
500 & 0.0032 & 0.0091 & 0.0077 & 0.0073 & 0.0034\\
1000 & 0.0021 & 0.0051 & 0.0047 & 0.0041 & 0.0020\\
5000 & 0.0002 & 0.0007 & 0.0007 & 0.0006 & 0.0003\\
\hline
&\multicolumn{5}{c}{AIMSE}\\
\hline
n&Oracle&MULTI-2SLS&PW-2SLS&SIMEX&Naive\\
\hline
100 & 0.0138 & 0.0465 & 0.0370 & 0.0394 & 0.0636\\
500 & 0.0051 & 0.0108 & 0.0096 & 0.0134 & 0.0518\\
1000 & 0.0025 & 0.0055 & 0.0050 & 0.0089 & 0.0496\\
5000 & 0.0003 & 0.0008 & 0.0007 & 0.0046 & 0.0470\\
\hline
\end{tabular}
\end{table}

\subsection{Simulation Study 2}
In the second simulation study, we examined the impact of measurement error distributions on the estimated functional coefficient $\widehat \beta_{1}(t)$ with a fixed sample size of $n = 1000$. The variance-covariance structure of the functional terms was restricted to $AR(1)$ with correlations of $\rho_{X} = \rho_{W} = \rho_{M} = 0.5$ and $\varepsilon_{X}(t) \sim MVN(0, \Sigma_{X})$ with $\Sigma_{X} = 1.5$. $M(t)$ was simulated as $M(t) = \delta(t)X(t) + \eta(t)$, where $\delta(t) = 0.5 \sin(2 \pi t)+1$ and $\eta(t) \sim MVN(0, \Sigma_{M})$ with $\Sigma_{M} = 1$. We considered the normal distribution, the $t$-distribution, and the Laplace distribution for measurement error, $U(t)$, with $\mu_{U}= 0$ and $\Sigma_{U} = 1$.

In table~\ref{table:me_dist}, the Oracle estimator consistently exhibits the lowest bias, variance, and AIMSE across all measurement error distributions. Conversely, the Naive estimator performs the poorest, with the highest AIMSE in all categories. The least squares-based methods (MULTI-2SLS and PW-2SLS) demonstrate similar performance. They have the same bias as the Oracle estimator, but their variances are higher than the Oracle estimator. Although SIMEX estimators are competitive, they tend to have higher bias but lower variance compared to least squares-based methods. The table also underscores the robustness of the five estimators when faced with different measurement error distributions.

\begin{table}[ht]
\centering\tiny\setlength\tabcolsep{3.5pt}
\caption{The influence of measurement error distributions (ME dist) for the functional variables on the performance of five estimators (Oracle, MULTI-2SLS, PW-2SLS, SIMEX, and Naive) in estimating $\beta_{1}(t)$, $n = 1000$. The functional variables were simulated under AR(1) covariance structures with $\rho_{X} = \rho_{W} = \rho_{M} = 0.5$, $\varepsilon_{X}(t) \sim MVN(0, \Sigma_{X})$ with $\Sigma_{X} = 1.5$, $\mu_{U}= 0$, $\Sigma_{U} = 1$, $\eta(t) \sim MVN(0, \Sigma_{M})$ with $\Sigma_{M} = 1$, $\delta(t) = 0.5  \sin(2 \pi t)+1$.}
\label{table:me_dist}
\centering
\begin{tabular}{r|ccccc}
\hline
&\multicolumn{5}{c}{$\text{ABias}^{2}$ (MSPEE, \%)}\\
\hline
ME dist&Oracle&MULTI-2SLS&PW-2SLS&SIMEX&Naive\\
\hline
Normal & 0.0004 (2.7) & 0.0003 (2.4) & 0.0004 (2.5) & 0.0048 (9.8) & 0.0476 (31.0)\\
t & 0.0006 (3.1) & 0.0005 (2.9) & 0.0004 (2.8) & 0.0049 (9.9) & 0.0475 (31.0)\\
Laplace & 0.0005 (3.0) & 0.0004 (2.7) & 0.0004 (2.7) & 0.0048 (9.8) & 0.0475 (31.0)\\
\hline
&\multicolumn{5}{c}{AVar}\\
\hline
ME dist&Oracle&MULTI-2SLS&PW-2SLS&SIMEX&Naive\\
\hline
Normal & 0.0021 & 0.0051 & 0.0047 & 0.0041 & 0.0020\\
t & 0.0021 & 0.0049 & 0.0046 & 0.0039 & 0.0020\\
Laplace & 0.0021 & 0.0051 & 0.0048 & 0.0041 & 0.0021\\
\hline
&\multicolumn{5}{c}{AIMSE}\\
\hline
ME dist&Oracle&MULTI-2SLS&PW-2SLS&SIMEX&Naive\\
\hline
Normal & 0.0025 & 0.0055 & 0.0050 & 0.0089 & 0.0496\\
t & 0.0026 & 0.0054 & 0.0050 & 0.0088 & 0.0495\\
Laplace & 0.0026 & 0.0056 & 0.0052 & 0.0089 & 0.0497\\
\hline
\end{tabular}
\end{table}

\subsection{Simulation Study 3}
In the third simulation study, we investigated the influence of the variance-covariance structure and its correlation for the functional error terms on the estimation of $\beta_{1}(t)$. $n = 1000$. The magnitudes of the functional error terms for $X(t)$, $W(t)$, and $M(t)$) were set as $\varepsilon_{X}(t) \sim MVN(0, \Sigma_{X})$ with $\Sigma_{X} = 1.5$; $U(t) \sim MVN(0, \Sigma_{U})$ with $\Sigma_{U} = 1$; and $\eta(t) \sim MVN(0, \Sigma_{M})$ with $\Sigma_{M} = 1$. The functional coefficient of $M(t)$, $\delta(t)$, was equal to $0.5 \sin(2 \pi t)+1$. We assumed that the variance-covariance structures of the functional terms are the same and considered independent (IND), $AR(1)$, compound symmetry (CS), and unstructured (UN) covariance structures with correlations of $\rho = (0.25, 0.5, 0.75)$. $\rho = 0$ under the IND covariance structure. 

\begin{figure}[ht]
\centering
\includegraphics[width=15cm,height=10cm]{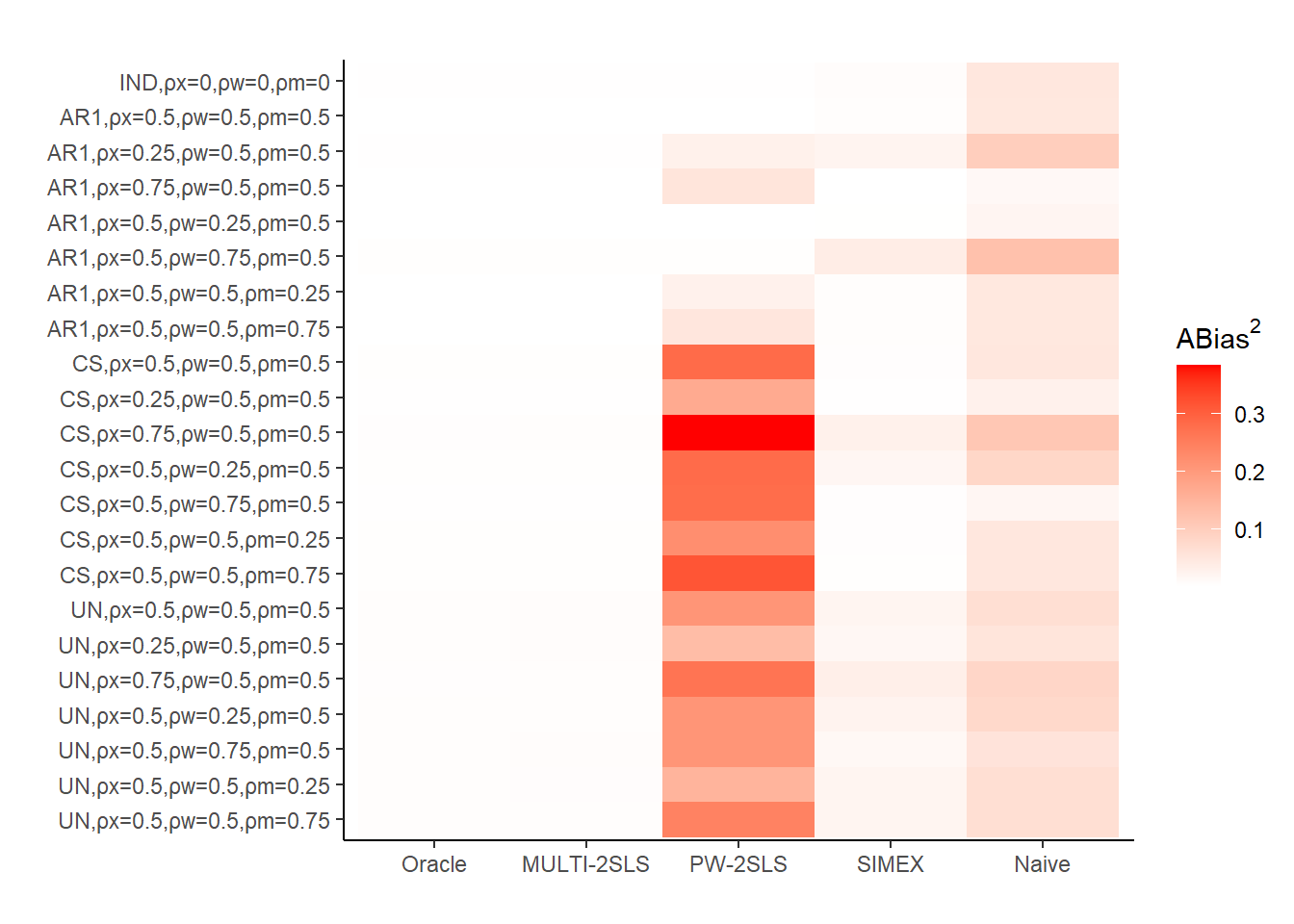}
\caption{The influence of the variance-covariance structures (Struct.) and their correlations ($\rho$) for the functional error terms on the $\text{ABias}^{2}$ of five estimators (Oracle, MULTI-2SLS, PW-2SLS, SIMEX, and Naive) in estimating $\beta_{1}(t)$, $n = 1000$, the functional variables with $\varepsilon_{X}(t) \sim MVN(0, \Sigma_{X})$ with $\Sigma_{X} = 1.5$, $U(t) \sim MVN(0, \Sigma_{U})$ with $\Sigma_{U} = 1$, $\eta(t) \sim MVN(0, \Sigma_{M})$ with $\Sigma_{M} = 1$, $\delta(t) = 0.5 \sin(2 \pi t)+1$.}
\label{sim3_bias}
\end{figure}

\begin{figure}[ht]
\centering
\includegraphics[width=15cm,height=10cm]{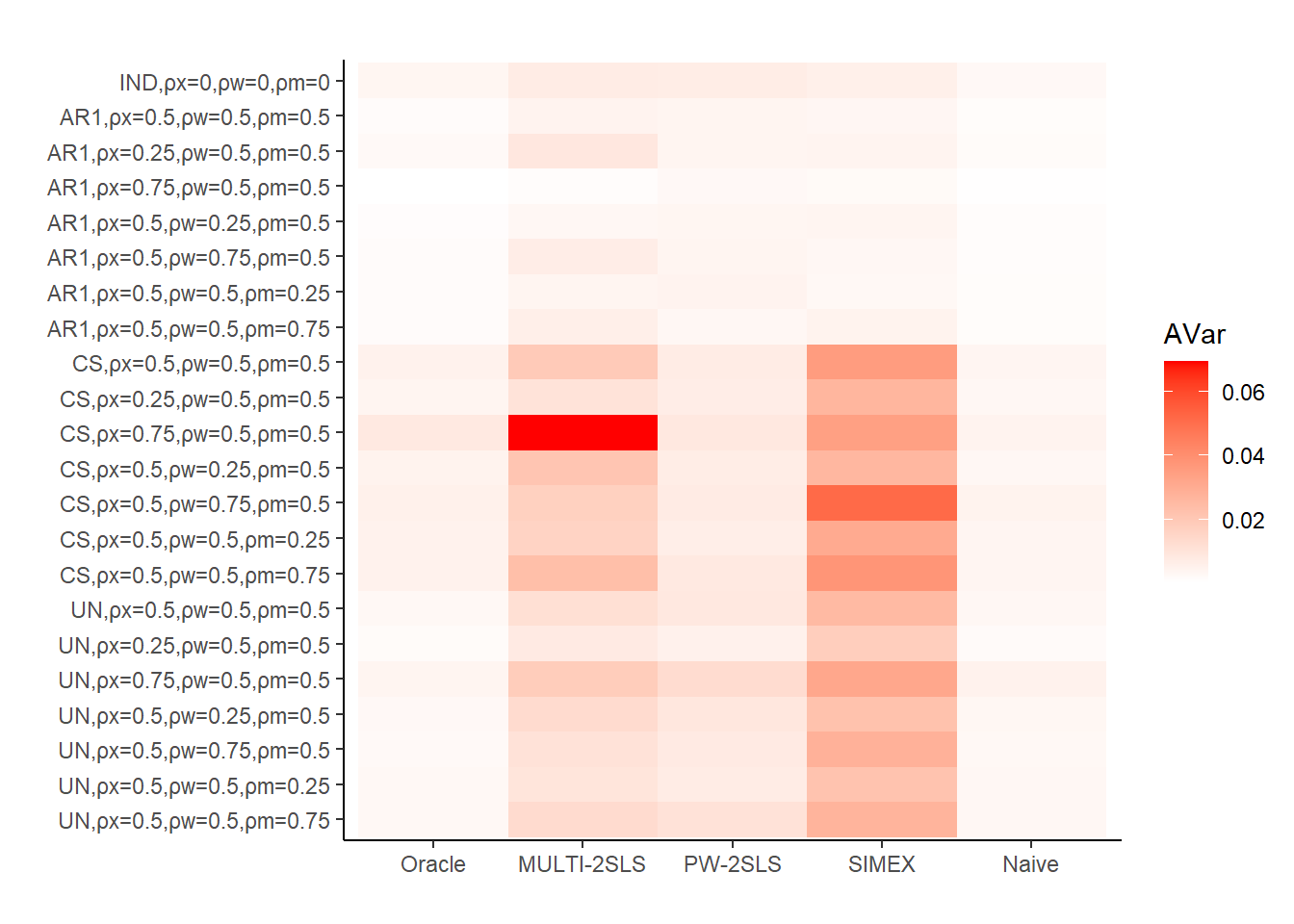}
\caption{The influence of the variance-covariance structures (Struct.) and their correlations ($\rho$) for the functional error terms on the $\text{AVar}$ of five estimators (Oracle, MULTI-2SLS, PW-2SLS, SIMEX, and Naive) in estimating $\beta_{1}(t)$, $n = 1000$, the functional variables with $\varepsilon_{X}(t) \sim MVN(0, \Sigma_{X})$ with $\Sigma_{X} = 1.5$, $U(t) \sim MVN(0, \Sigma_{U})$ with $\Sigma_{U} = 1$, $\eta(t) \sim MVN(0, \Sigma_{M})$ with $\Sigma_{M} = 1$, $\delta(t) = 0.5 \sin(2 \pi t)+1$.}
\label{sim3_var}
\end{figure}

\begin{figure}[ht]
\centering
\includegraphics[width=15cm,height=10cm]{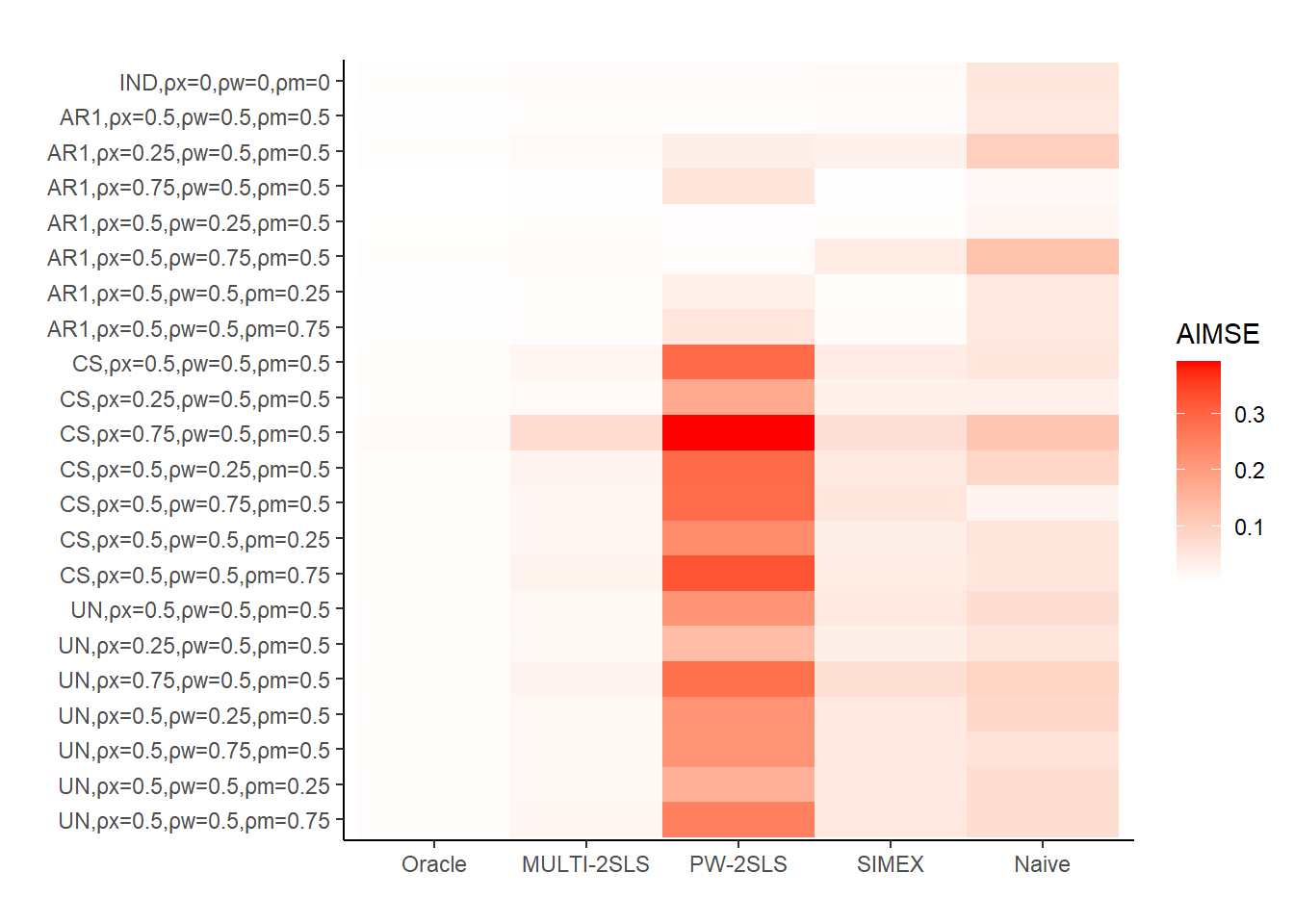}
\caption{The influence of the variance-covariance structures (Struct.) and their correlations ($\rho$) for the functional error terms on the AISME of five estimators (Oracle, MULTI-2SLS, PW-2SLS, SIMEX, and Naive) in estimating $\beta_{1}(t)$, $n = 1000$, the functional variables with $\varepsilon_{X}(t) \sim MVN(0, \Sigma_{X})$ with $\Sigma_{X} = 1.5$, $U(t) \sim MVN(0, \Sigma_{U})$ with $\Sigma_{U} = 1$, $\eta(t) \sim MVN(0, \Sigma_{M})$ with $\Sigma_{M} = 1$, $\delta(t) = 0.5 \sin(2 \pi t)+1$.}
\label{sim3_aimse}
\end{figure}

Figures~\ref{sim3_bias}, ~\ref{sim3_var}, and ~\ref{sim3_aimse} show that the Oracle estimator outperforms the other methods on all metrics, making it the most reliable method. The PW-2SLS method was highly sensitive to changes in the variance-covariance structure and the correlations of functional terms. The estimator performed well only under the IND and $AR(1)$ structures, particularly when $\rho_{X} = \rho_{M}$, while being less affected by variations in $\rho_{W}$. Under the CS and UN structures, the estimators showed substantially larger bias, with the bias increasing as $\rho_{X}$ and $\rho_{M}$ increased. The performance of the MULTI-2SLS and SIMEX methods, in terms of bias, variance, and AIMSE, remained relatively stable across various covariance structures and correlation settings. In conclusion, the choice of the most suitable method depends on the underlying data structure and correlations. Although the Oracle method is optimal, it is impractical to implement in real-world applications because the true covariate $X(t)$ is latent. In contrast, the MULTI-2SLS method offers a more feasible alternative, with low bias and competitive variance, making it a viable and reasonable choice. The detailed simulation results are available in the Appendix A.

\subsection{Simulation Study 4}
In the fourth simulation study, we evaluated the impact of the magnitudes of the functional predictor on the estimation of the functional coefficient $\beta_{1}(t)$. The sample size was set at $n = 1000$. The covariance structure of the functional terms was AR(1) with $\rho_{X} = \rho_{W} = \rho_{M} = 0.5$. The instrumental variable was simulated as $M(t) = \delta(t)X(t) + \eta(t)$, where $\delta(t) = 0.5 \sin(2 \pi t)+1$ and $\eta(t) \sim MVN(0, \Sigma_{M})$ with $\Sigma_{M} = 1$. We considered the magnitudes of $\varepsilon_{X}(t) \sim MVN(0, \Sigma_{X})$ with $\Sigma_{X}=(1, 1.5, 2, 4)$, $U(t) \sim MVN(0, \Sigma_{U})$ with $\Sigma_{U}=(0.5, 1, 2)$, and $Ratio=\Sigma_{X}/\Sigma_{U}$.

Table~\ref{table:FVSD} shows that, as the ratio of $\Sigma_{X}$ and $\Sigma_{U}$ increases, the biases of all estimators decrease. The Oracle, MULTI-2SLS, and PW-2SLS estimators consistently exhibited lower bias compared to the other two estimators across all ratios, indicating that they were minimally affected by changes in the magnitudes of the true functional predictor and its measurement error. In contrast, the SIMEX and Naive estimators showed higher biases, especially when the ratio was small (ratio $<$ 2). In particular, the MULTI-2SLS estimators also had relatively higher variances compared to the other estimators in cases where the ratio was small (ratio $<$ 2). Overall, when considering both bias and variance (i.e., AIMSE), the Oracle estimator demonstrated the best performance, followed by the MULTI-2SLS, PW-2SLS, SIMEX, and Naive estimators.

\begin{table}[ht]
\centering\tiny\setlength\tabcolsep{3.5pt}
\caption{The influence of standard deviations of the true functional predictor and the measurement error on the performance of five estimators (Oracle, MULTI-2SLS, PW-2SLS, SIMEX, and Naive) in estimating $\beta_{1}(t)$, $n = 1000$. The functional variables were simulated with AR(1) structure with$\rho_{X} = \rho_{W} = \rho_{M} = 0.5$, $\varepsilon_{X}(t) \sim MVN(0, \Sigma_{X})$, $U(t) \sim MVN(0, \Sigma_{U})$, $\eta(t) \sim MVN(0, \Sigma_{M})$ with $\Sigma_{M} = 1$, $\delta(t) = 0.5 \sin(2 \pi t)+1$.}
\label{table:FVSD}
\centering
\begin{tabular}{rrr|ccccc}
\hline
&&&\multicolumn{5}{c}{$\text{ABias}^{2}$ (MSPEE, \%)}\\
\hline
$\Sigma_{X}$&$\Sigma_{U}$&Ratio&Oracle&MULTI-2SLS&PW-2SLS&SIMEX&Naive\\
\hline
1.0 & 2.0 & 0.50 & 0.0040 (8.6) & 0.0041 (8.7) & 0.0040 (8.7) & 0.2246 (67.3) & 0.3184 (80.2)\\
1.5 & 2.0 & 0.75 & 0.0030 (7.4) & 0.0027 (7.1) & 0.0030 (7.4) & 0.0964 (44.1) & 0.2040 (64.2)\\
1.0 & 1.0 & 1.00 & 0.0027 (7.1) & 0.0027 (7.0) & 0.0026 (6.9) & 0.0367 (27.2) & 0.1248 (50.2)\\
2.0 & 2.0 & 1.00 & 0.0013 (4.8) & 0.0012 (4.6) & 0.0013 (4.8) & 0.0360 (27.0) & 0.1248 (50.2)\\
1.5 & 1.0 & 1.50 & 0.0004 (2.7) & 0.0003 (2.4) & 0.0004 (2.5) & 0.0048 ( 9.8) & 0.0476 (31.0)\\
1.0 & 0.5 & 2.00 & 0.0009 (4.0) & 0.0009 (4.0) & 0.0008 (3.9) & 0.0015 ( 5.3) & 0.0206 (20.4)\\
2.0 & 1.0 & 2.00 & 0.0001 (1.2) & 0.0001 (1.1) & 0.0001 (1.2) & 0.0008 ( 4.0) & 0.0200 (20.1)\\
4.0 & 2.0 & 2.00 & 0.0001 (1.1) & 0.0001 (1.1) & 0.0001 (1.1) & 0.0006 ( 3.5) & 0.0199 (20.1)\\
1.5 & 0.5 & 3.00 & 0.0001 (1.2) & 0.0001 (1.2) & 0.0001 (1.3) & 0.0001 ( 1.2) & 0.0051 (10.2)\\
2.0 & 0.5 & 4.00 & 0.0001 (1.1) & 0.0001 (1.1) & 0.0001 (1.1) & 0.0001 ( 1.2) & 0.0018 ( 6.1)\\
4.0 & 1.0 & 4.00 & 0.0001 (1.0) & 0.0001 (1.0) & 0.0001 (1.0) & 0.0001 ( 1.1) & 0.0018 ( 6.0)\\
4.0 & 0.5 & 8.00 & 0.0000 (0.9) & 0.0000 (0.9) & 0.0000 (0.9) & 0.0001 ( 1.1) & 0.0002 ( 1.8)\\
\hline
&&&\multicolumn{5}{c}{AVar}\\
\hline
$\Sigma_{X}$&$\Sigma_{U}$&Ratio&Oracle&MULTI-2SLS&PW-2SLS&SIMEX&Naive\\
\hline
1.0 & 2.0 & 0.50 & 0.0024 & 0.0343 & 0.0088 & 0.0025 & 0.0009\\
1.5 & 2.0 & 0.75 & 0.0018 & 0.0109 & 0.0039 & 0.0030 & 0.0013\\
1.0 & 1.0 & 1.00 & 0.0030 & 0.0127 & 0.0096 & 0.0051 & 0.0023\\
2.0 & 2.0 & 1.00 & 0.0018 & 0.0063 & 0.0031 & 0.0029 & 0.0015\\
1.5 & 1.0 & 1.50 & 0.0021 & 0.0051 & 0.0047 & 0.0041 & 0.0020\\
1.0 & 0.5 & 2.00 & 0.0035 & 0.0080 & 0.0098 & 0.0081 & 0.0032\\
2.0 & 1.0 & 2.00 & 0.0008 & 0.0021 & 0.0022 & 0.0027 & 0.0012\\
4.0 & 2.0 & 2.00 & 0.0002 & 0.0011 & 0.0005 & 0.0012 & 0.0007\\
1.5 & 0.5 & 3.00 & 0.0013 & 0.0025 & 0.0039 & 0.0042 & 0.0014\\
2.0 & 0.5 & 4.00 & 0.0006 & 0.0011 & 0.0020 & 0.0023 & 0.0007\\
4.0 & 1.0 & 4.00 & 0.0002 & 0.0004 & 0.0005 & 0.0009 & 0.0003\\
4.0 & 0.5 & 8.00 & 0.0002 & 0.0002 & 0.0006 & 0.0006 & 0.0002\\
\hline
&&&\multicolumn{5}{c}{AIMSE}\\
\hline
$\Sigma_{X}$&$\Sigma_{U}$&Ratio&Oracle&MULTI-2SLS&PW-2SLS&SIMEX&Naive\\
\hline
1.0 & 2.0 & 0.50 & 0.0064 & 0.0384 & 0.0128 & 0.2271 & 0.3194\\
1.5 & 2.0 & 0.75 & 0.0048 & 0.0136 & 0.0069 & 0.0994 & 0.2053\\
1.0 & 1.0 & 1.00 & 0.0057 & 0.0153 & 0.0121 & 0.0418 & 0.1271\\
2.0 & 2.0 & 1.00 & 0.0031 & 0.0075 & 0.0044 & 0.0389 & 0.1263\\
1.5 & 1.0 & 1.50 & 0.0025 & 0.0055 & 0.0050 & 0.0089 & 0.0496\\
1.0 & 0.5 & 2.00 & 0.0044 & 0.0089 & 0.0107 & 0.0096 & 0.0238\\
2.0 & 1.0 & 2.00 & 0.0009 & 0.0022 & 0.0023 & 0.0035 & 0.0212\\
4.0 & 2.0 & 2.00 & 0.0002 & 0.0012 & 0.0006 & 0.0018 & 0.0206\\
1.5 & 0.5 & 3.00 & 0.0013 & 0.0025 & 0.0040 & 0.0043 & 0.0066\\
2.0 & 0.5 & 4.00 & 0.0006 & 0.0012 & 0.0020 & 0.0024 & 0.0026\\
4.0 & 1.0 & 4.00 & 0.0002 & 0.0005 & 0.0006 & 0.0010 & 0.0021\\
4.0 & 0.5 & 8.00 & 0.0002 & 0.0003 & 0.0006 & 0.0006 & 0.0004\\
\hline
\end{tabular}
\end{table}

\subsection{Simulation Study 5}
In the fifth simulation study, we evaluated the impacts of the magnitudes of the standard deviation and $\delta(t)$ associated with the instrumental variable in estimating the functional coefficient with a sample size of $n = 1000$. The variance-covariance structure of functional-valued variables (i.e., $X(t)$, $W(t)$, and $M(t)$) was restricted to $AR(1)$ with correlations of $\rho_{X} = \rho_{W} = \rho_{M} = 0.5$ and $\varepsilon_{X}(t) \sim MVN(0, \Sigma_{X})$ with $\Sigma_{X} = 1.5$, $U(t) \sim MVN(0, \Sigma_{U})$ with $\Sigma_{U} = 1$. We simulated $\eta(t)$ as $\eta(t) \sim MVN(0, \Sigma_{M})$ with $\Sigma_{M}=(0.5, 1, 2, 4)$, and $\delta(t) = c \sin(2 \pi t)+1$ with $c=(0, 0.25, 0.5, 0.75)$ for the instrumental variable $\{M(t)\}$. When evaluating the impacts of varying $\Sigma_{M}$, $c$ was kept constant at $c=0.50$. Similarly, $\Sigma_{M}$ was maintained constant at 1 when assessing the impacts of varying $c$ on the estimation under the five different estimators considered in our simulation studies. The performance of Oracle and Naive estimators was not influenced by the changing magnitudes of the instrumental variable. 

In Tables~\ref{table:IVSD} and ~\ref{table:IVDelta}, we observed that the performance metrics (bias, variance, and AIMSE) of estimators with measurement error adjustment deteriorated as standard deviations and the value of $c$ for the instrumental variable increased. However, the extent of this decline varied between methods. Across all comparisons among the four measurement error correction approaches, the MULTI-2SLS and PW-2SLS methods performed the best in terms of biases and AIMSEs. Alternatively, the SIMEX estimator had slightly smaller variances.

\begin{table}[ht]
\centering\tiny\setlength\tabcolsep{3.5pt}
\caption{The impacts of the magnitudes of the instrumental variable in estimating $\beta_{1}(t)$ with $n = 1000$. The functional variables were simulated under AR(1) covariance structures with $\rho_{X} = \rho_{W} = \rho_{M} = 0.5$, $\varepsilon_{X}(t) \sim MVN(0, \Sigma_{X})$ with $\Sigma_{X} = 1.5$, $U(t) \sim MVN(0, \Sigma_{U})$ with $\Sigma_{U} = 1$, $\eta(t) \sim MVN(0, \Sigma_{M})$, $\delta(t) = 0.5 \sin(2 \pi t)+1$.}
\label{table:IVSD}
\centering
\begin{tabular}{r|ccccc}
\hline
&\multicolumn{5}{c}{$\text{ABias}^{2}$ (MSPEE, \%)}\\
\hline
$\Sigma_{M}$&Oracle&MULTI-2SLS&PW-2SLS&SIMEX&Naive\\
\hline
0.5 & 0.0004 (2.7) & 0.0003 (2.4) & 0.0004 (2.6) & 0.0047 ( 9.7) & 0.0476 (31.0)\\
1.0 & 0.0004 (2.7) & 0.0003 (2.4) & 0.0004 (2.5) & 0.0048 ( 9.8) & 0.0476 (31.0)\\
2.0 & 0.0004 (2.7) & 0.0003 (2.4) & 0.0003 (2.5) & 0.0054 (10.4) & 0.0476 (31.0)\\
4.0 & 0.0004 (2.7) & 0.0004 (2.5) & 0.0004 (2.6) & 0.0065 (11.4) & 0.0476 (31.0)\\
\hline
&\multicolumn{5}{c}{AVar}\\
\hline
$\Sigma_{M}$&Oracle&MULTI-2SLS&PW-2SLS&SIMEX&Naive\\
\hline
0.5 & 0.0021 & 0.0041 & 0.0028 & 0.0033 & 0.0020\\
1.0 & 0.0021 & 0.0051 & 0.0047 & 0.0041 & 0.0020\\
2.0 & 0.0021 & 0.0095 & 0.0114 & 0.0070 & 0.0020\\
4.0 & 0.0021 & 0.0297 & 0.0382 & 0.0152 & 0.0020\\
\hline
&\multicolumn{5}{c}{AIMSE}\\
\hline
$\Sigma_{M}$&Oracle&MULTI-2SLS&PW-2SLS&SIMEX&Naive\\
\hline
0.5 & 0.0025 & 0.0044 & 0.0032 & 0.0080 & 0.0496\\
1.0 & 0.0025 & 0.0055 & 0.0050 & 0.0089 & 0.0496\\
2.0 & 0.0025 & 0.0098 & 0.0118 & 0.0124 & 0.0496\\
4.0 & 0.0025 & 0.0301 & 0.0386 & 0.0217 & 0.0496\\
\hline
\end{tabular}
\end{table}

\begin{table}[ht]
\centering\tiny\setlength\tabcolsep{3.5pt}
\caption{The impacts of the magnitudes of the instrumental variable in estimating $\beta_{1}(t)$ with $n = 1000$. The functional variables were simulated under AR(1) covariance structures with $\rho_{X} = \rho_{W} = \rho_{M} = 0.5$, $\varepsilon_{X}(t) \sim MVN(0, \Sigma_{X})$ with $\Sigma_{X} = 1.5$, $U(t) \sim MVN(0, \Sigma_{U})$ with $\Sigma_{U} = 1$, $\eta(t) \sim MVN(0, \Sigma_{M})$ with $\Sigma_{M} = 1$, $\delta(t) = c \sin(2 \pi t)+1$.}
\label{table:IVDelta}
\centering
\begin{tabular}{r|ccccc}
\hline
&\multicolumn{5}{c}{$\text{ABias}^{2}$ (MSPEE, \%)}\\
\hline
c&Oracle&MULTI-2SLS&PW-2SLS&SIMEX&Naive\\
\hline
0.00 & 0.0004 (2.7) & 0.0003 (2.4) & 0.0004 (2.6) & 0.0048 ( 9.8) & 0.0476 (31.0)\\
0.25 & 0.0004 (2.7) & 0.0003 (2.4) & 0.0004 (2.6) & 0.0048 ( 9.8) & 0.0476 (31.0)\\
0.50 & 0.0004 (2.7) & 0.0003 (2.4) & 0.0004 (2.5) & 0.0048 ( 9.8) & 0.0476 (31.0)\\
0.75 & 0.0004 (2.7) & 0.0003 (2.3) & 0.0003 (2.5) & 0.0051 (10.1) & 0.0476 (31.0)\\
\hline
&\multicolumn{5}{c}{AVar}\\
\hline
c&Oracle&MULTI-2SLS&PW-2SLS&SIMEX&Naive\\
\hline
0.00 & 0.0021 & 0.0047 & 0.0040 & 0.0035 & 0.0020\\
0.25 & 0.0021 & 0.0048 & 0.0041 & 0.0037 & 0.0020\\
0.50 & 0.0021 & 0.0051 & 0.0047 & 0.0041 & 0.0020\\
0.75 & 0.0021 & 0.0062 & 0.0061 & 0.0056 & 0.0020\\
\hline
&\multicolumn{5}{c}{AIMSE}\\
\hline
c&Oracle&MULTI-2SLS&PW-2SLS&SIMEX&Naive\\
\hline
0.00 & 0.0025 & 0.0050 & 0.0044 & 0.0083 & 0.0496\\
0.25 & 0.0025 & 0.0051 & 0.0045 & 0.0085 & 0.0496\\
0.50 & 0.0025 & 0.0055 & 0.0050 & 0.0089 & 0.0496\\
0.75 & 0.0025 & 0.0065 & 0.0064 & 0.0107 & 0.0496\\
\hline
\end{tabular}
\end{table}

\section{Application}\label{sec:application}

To assess the impact of physical activity on body mass index (BMI), we applied our methods to the NHANES data 2003-2004 and 2005-2006 cycles \cite{nhanes_home}. NHANES tracked participants' daily physical activity levels using a physical activity monitor (PAM) from Pensacola, Florida, which recorded uniaxial movement and logged the intensity of acceleration every minute \cite{nhanes2005paxraw}. Participants were instructed to wear the PAM at home for seven consecutive days during waking hours. The device was not worn during sleep or any water-based activities. We summed up data on the intensity of physical activity at the minute level into measures at the hour level and excluded extreme physical activity intensity values, specifically those exceeding three times the interquartile range above the third quartile. For each participant, we only included data from days with complete 24-hour recordings. Participants were included in the analysis if they were at least 20 years old and had physical activity data for at least one weekday and one weekend day. And we calculated weekday and weekend physical activity intensity by averaging the values across all recorded weekdays and weekends, respectively.

A functional linear regression model was applied to evaluate the association between weekday physical activity intensity and BMI. In our application, BMI was the outcome, and device-based weekday physical intensity and weekend physical intensity data were considered to be the functional exposure prone to measurement error and the functional instrumental variable, respectively. Weekend physical activity intensity is chosen as an instrumental variable for weekday intensity based on the rationale that it is strongly correlated with individuals’ true weekday activity levels, yet retains enough variation. Activity patterns often differ between weekdays and weekends due to changes in daily routines, available free time, and social environments. In addition, we included the age, gender, race/ethnicity, and diabetes status as error-free covariates in the model. Sample weights were adjusted in all analyses according to the NHANES analytic guidelines \cite{johnson2013national}. The 95\% confidence interval for the estimators was obtained using the nonparametric bootstrap over $500$ replicates.

\subsection{Final Analytic Data}
\begin{figure}[ht]
\centering
\includegraphics[width=10cm,height=10cm]{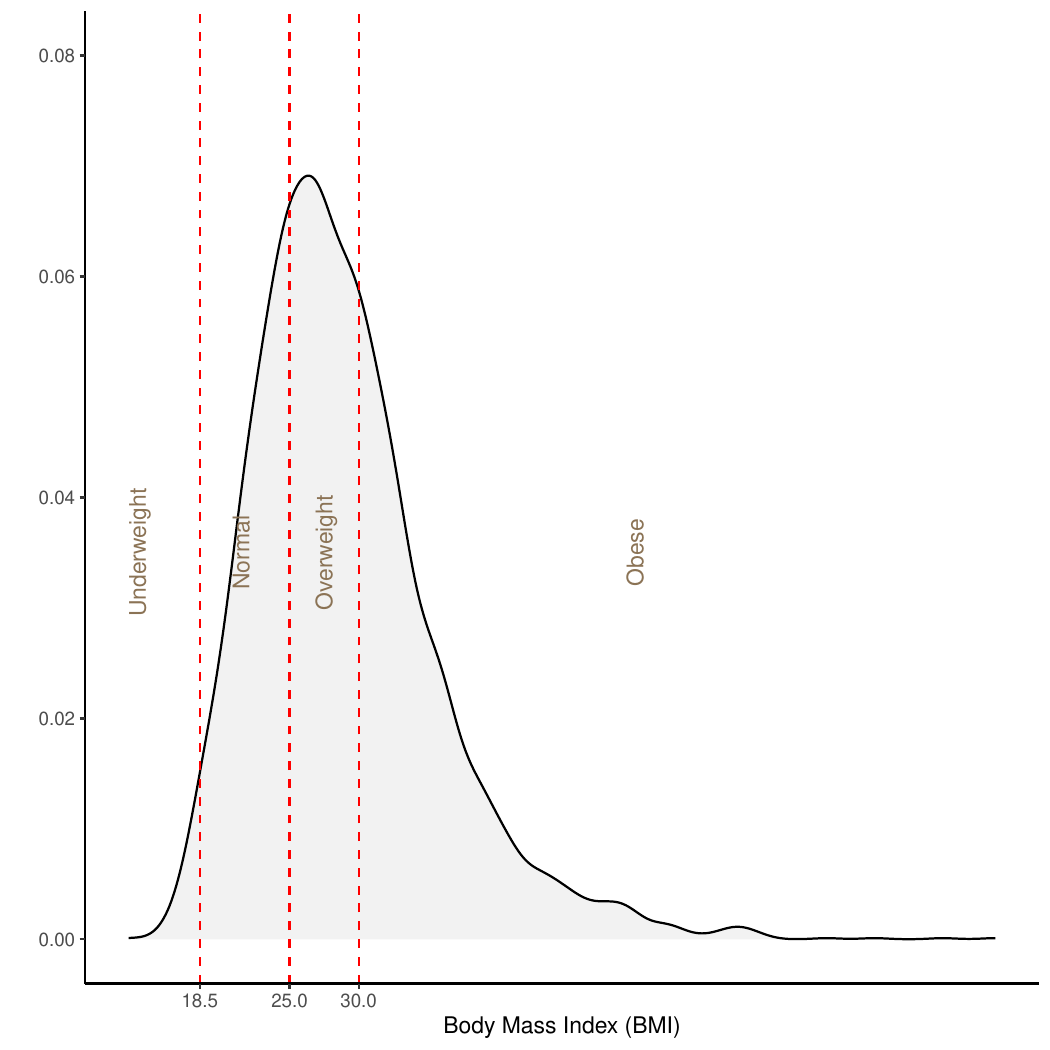}
\caption{Density plot of BMI.}
\label{bmi_distribution}
\end{figure}

\begin{figure}[ht]
\centering
\includegraphics[width=17cm,height=8cm]{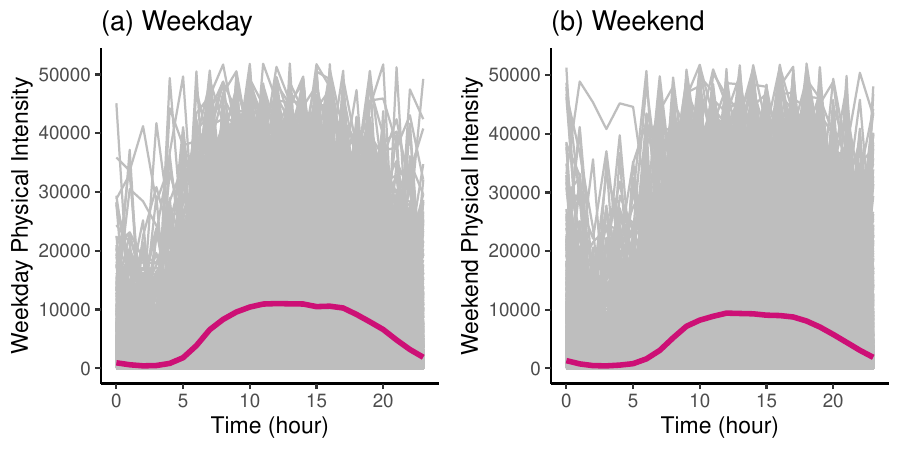}
\caption{Plots of device-based weekday and weekend physical intensity over time. Grey lines are individual-specific trajectories, and the pink line shows the overall trajectory averaged across participants.}
\label{pa_distribution}
\end{figure}

The final analytic weighted sample used in our application included 153,792,444 community-dwelling adults between the ages of 20 to 85 living in the United States, including $55.0\%$ females. Participants' average age was $47.1  (\pm SD=17.0)$ years, and $71.8\%$ were white, $12.3\%$ were black, $10.9\%$ were Hispanic, and $4.9\%$ were of another race/ethnicity. Participants' average BMI was $28.9 (\pm SD=6.8)$ kg/m$^2$. We summarize the demographic characteristics of the adults included in the analysis in Table~\ref{table:demo}. The distribution of the outcome, BMI, is shown in Figure~\ref{bmi_distribution}. Figure~\ref{pa_distribution} presents the distribution of weekday and weekend physical activity intensity against wear time for the participants; the patterns were similar during weekdays and weekends. Over 40\% of the physical intensity values during hours 1–6 and 23 were zeros, which largely correspond to typical sleep hours. Therefore, we excluded these hours from the following analyses.

\begin{table}[ht]
\centering\small\setlength\tabcolsep{4pt}
\caption{Demographics of the study population.}
\label{table:demo}
\centering
\begin{tabular}{rl|c}
\hline
&&Overall\\
\hline
N&&153792444\\
Age (year, mean (SD))&&47.1 (17.0)\\
Gender (\%)&Male&69263428 (45.0)\\
&Female&84529016 (55.0)\\
Race/Ethnicity (\%)&Non-Hispanic White&110479083 (71.8)\\
&Non-Hispanic Black&18972669 (12.3)\\
&Hispanic&16838600 (10.9)\\
&Other Race - Including Multi-Racial&7502093 (4.9)\\
Education (\%)&High School Grad/GED or Lower&67661851 (44.1)\\
&Some College or AA degree&48258283 (31.4)\\
&College Graduate or above&37667493 (24.5)\\
Diabetes (\%)&No&140467096 (91.3)\\
&Yes&13325349 (8.7)\\
BMI (kg/m2, mean (SD))&&28.9 (6.8)\\
\hline
\end{tabular}
\end{table}

\subsection{Data Application Results}
\begin{figure}[ht]
\centering
\includegraphics[width=17cm,height=16cm]{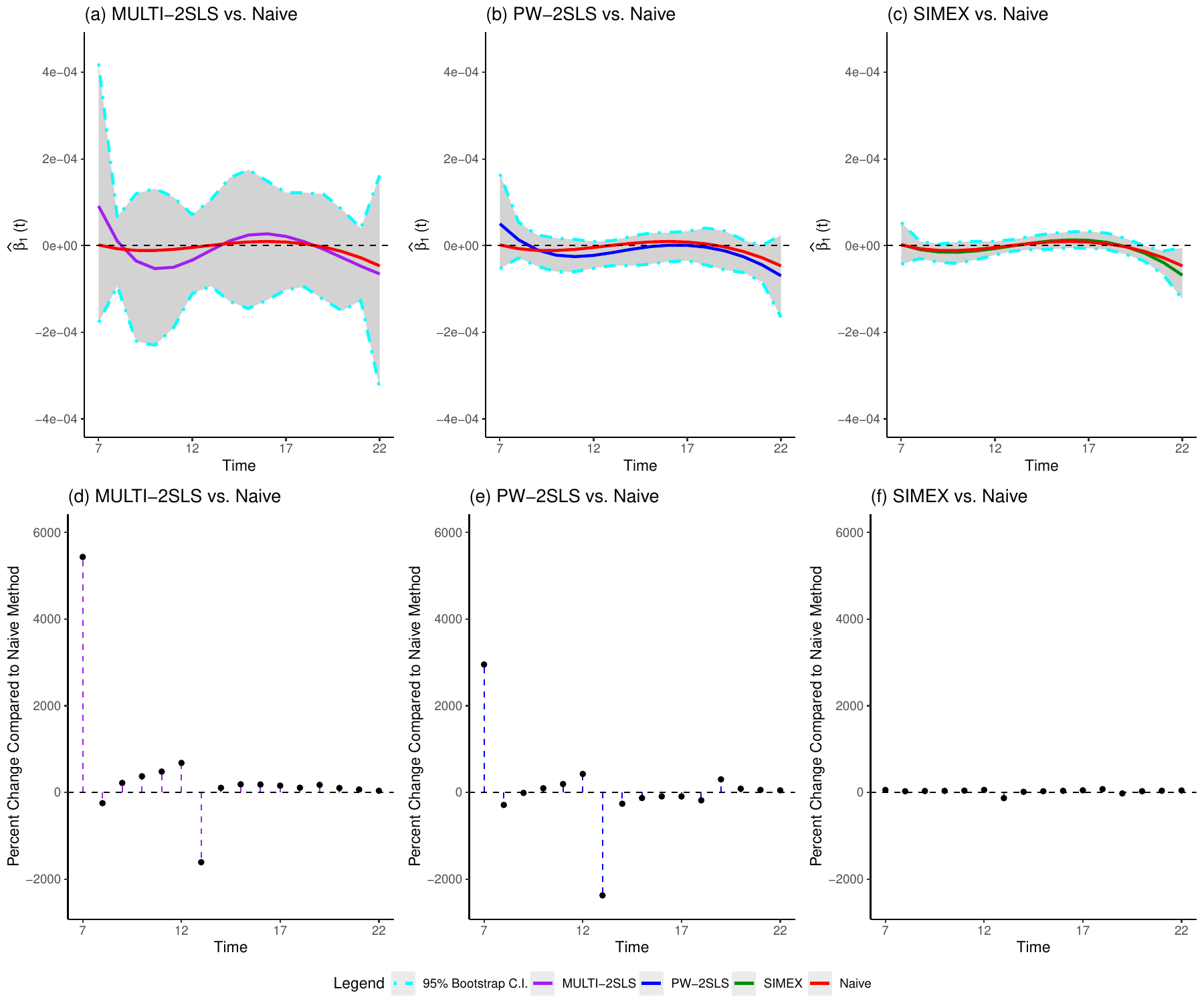}
\caption{(a)-(c): Plots of functional estimators by time $\{\widehat\beta_{1}(t)\}$ and their 95\% nonparametric bootstrap confidence intervals. Portions of the grey-shaded area for confidence intervals that are either completely above or below the dotted black line are statistically significant at the $5\%$ significance level for that time period. (d)-(f): Plots of pointwise percent change compared to the Naive method on functional estimators.}
\label{pa_beta1b}
\end{figure}

The estimated functional coefficients for the weekday physical intensity over time are illustrated in Figure~\ref{pa_beta1b} (a)-(c). The association between weekday physical intensity and BMI varied by time of the day, and the curves in the plot demonstrate moderate fluctuations and peak at around 7 a.m. and 4 p.m. MULTI-2SLS, PW-2SLS, and SIMEX methods yield estimates that differ from the Naive approach, with the estimators that adjusted for measurement error exhibiting greater fluctuations over time compared to the Naive method. Based on their 95\% bootstrap confidence intervals, weekday physical intensity showed a significant association with BMI during the end of the day (8 p.m., 9 p.m., and 10 p.m.) for the SIMEX estimator. In contrast, the Naive estimator only exhibited significant associations at 9 p.m. and 10 p.m., while no significant associations were found using the MULTI-2SLS and PW-2SLS approaches.

We compared the estimators obtained after correction for measurement errors (MULTI-2SLS, PW-2SLS, and SIMEX) with the Naive estimator because it was considered the least accurate method to estimate $\beta_{1}(t)$ based on the results of simulation studies (Figure~\ref{pa_beta1b} (d)-(f)). We also calculated the average percent difference between the measurement error-corrected estimators and the Naive estimator as follows: 
\begin{equation*}
\text{Percent Difference} = \text{mean} \left[ \Big\vert {\frac{\beta_{1,with\ ME\ correction}(t)-\beta_{1,Naive}(t)}{\beta_{1,Naive}(t)}} \Big\vert \times 100 \right]
\end{equation*}
The estimator of the MULTI-2SLS approach was the furthest from the Naive estimator, with a large percent difference of 636\% but with the widest confidence interval of $\widehat\beta_{1}(t)$, while the percent differences of PW-2SLS and SIMEX estimators were 475\% and 46\%, respectively, with narrower confidence intervals.

Table~\ref{table:application_EF} presents the coefficients of error-free covariates on BMI and their 95\% confidence intervals. The effects of age and gender on BMI were not statistically significant. Non-Hispanic Black participants had a significantly higher BMI compared to Non-Hispanic White participants, while participants of other racial groups had notably lower BMI than Non-Hispanic White participants. Participants with diabetes had a significantly higher BMI compared to those without diabetes. The results remained consistent across all estimators.

\begin{table}[t]
\centering\tiny\setlength\tabcolsep{1pt}
\caption{Estimated associations between error-free covariates and BMI under the four approaches.}
\label{table:application_EF}
\centering
\begin{tabular}{rc|cc|cc|cc|cc}
\hline
Error-free Covariates&&MULTI-2SLS&95\% C.I.&PW-2SLS&95\% C.I.&SIMEX&95\% C.I.&Naive&95\% C.I.\\
\hline
Age&&0.0003&(-0.01, 0.01)&0.0003&(-0.01, 0.01)&0.0014&(-0.01, 0.01)&0.0015&(-0.01, 0.02)\\
Gender&Male&Ref.&&&&&&&\\
&Female&0.09&(-0.46, 0.61)&0.09&(-0.45, 0.61)&0.03&(-0.52, 0.56)&0.03&(-0.51, 0.57)\\
Race/Ethnicity&Non-Hispanic White&Ref.&&&&&&&\\
&Non-Hispanic Black&1.64&(1.00, 2.34)&1.64&(1.00, 2.34)&1.71&(1.10, 2.42)&1.71&(1.10, 2.40)\\
&Hispanic&0.04&(-0.56, 0.73)&0.04&(-0.56, 0.73)&0.07&(-0.54, 0.74)&0.06&(-0.54, 0.75)\\
&Other Race - Including Multi-Racial&-2.15&(-3.42, -0.92)&-2.15&(-3.42, -0.91)&-2.10&(-3.38, -0.87)&-2.13&(-3.38, -0.89)\\
Diabetes&No&Ref.&&&&&&&\\
&Yes&3.95&(2.87, 5.02)&3.94&(2.88, 5.02)&3.95&(2.88, 5.06)&3.96&(2.89, 5.06)\\
\hline
\end{tabular}
\end{table}

Overall, the application results illustrate an obvious difference between the estimators with measurement error adjustments and the Naive estimator for the functional exposure. In general, not accounting for measurement errors in exposures or covariates can introduce bias by overestimating or underestimating exposure effects, leading to misrepresentations of the true relationships. By incorporating measurement error adjustments, estimates are generally more reliable and robust, especially in research fields where precise measurement is challenging.

\section{Discussion}\label{sec:discussion}

We proposed two two-stage least squares-based measurement error correction approaches with an instrumental variable for functional linear regression models. The methods are applicable for a continuous scalar-valued outcome with a function-valued covariate prone to error and error-free scalar-valued covariates. To adjust for measurement error bias, we implemented three measurement error correction approaches. Among the three approaches, the MULTI-2SLS method showed the strongest performance, consistently achieving a smaller bias in most simulation scenarios and lower AIMSE values. Additionally, it significantly outperformed the SIMEX method in terms of runtime (0.008 vs. 4.670 seconds for N=1,000 per simulation), making it the most efficient and scalable option overall. However, the SIMEX method exhibited slightly lower overall variance. In studies of behavioral interventions that involve accelerometer-based monitoring of physical activity, the resulting data are correlated functional data where the same subjects are continuously monitored over a period of time. This study design inherently introduces within-person correlations because physical activity measurements obtained from the same person are more likely to be similar than those obtained from other subjects. Thus, considering within-person correlation when correcting measurement errors in the functional-valued covariate resulted in optimized performance, as demonstrated by the simulation results of the MULTI-2SLS approach. In the application, the comparison of these methods demonstrates that the choice of method can lead to substantially different results from the same dataset, which underscores the importance of considering model assumptions, the structure of data, and potential measurement errors in statistical analyses.

Our least squares-based methods have some limitations. Regression calibration-based approaches have been demonstrated to be consistent and effectively adjust for measurement error biases in linear regression models. However, the use of regression calibration in nonlinear models results in approximate consistent estimates. Nonlinear models involve complex relationships between variables that are not easily adjusted through calibration alone. As a result, in nonlinear contexts, regression calibration typically serves as an approximation rather than a precise correction method. \cite{boe2023issues,carroll2006measurement}. On the other hand, the SIMEX method has been widely used to address measurement error in nonlinear regression models \cite{mao2017simex,shang2012measurement,sevilimedu2022simulation}. Our current measurement error adjustment methods are limited to handling two-dimensional functional predictors that are susceptible to measurement error. As multi-dimensional functional physical activity data becomes more widely available, we may explore extending these methods to accommodate more complex functional data structures in the future.

\bmsection*{Author contributions}
XC performed simulations and data analyses for the project and drafted the manuscript. LX proposed the MULTI-2SLS method and edited the draft. UB assisted with establishing the theoretical properties of the proposed methods and edited the draft. CQ, HJ, GH, and RSZ edited the draft. CDT conceived and supervised the project and edited the draft. All authors reviewed the manuscript.

\bmsection*{Acknowledgment}
This research was supported by awards from the National Institutes of Diabetes, Digestive, and Kidney Disease Award numbers R01DK132385 and 1R01DK136994-01A1. This research was also supported in part by Lilly Endowment, Inc., through its support for the Indiana University Pervasive Technology Institute.

\bmsection*{Conflict of interest}
The authors have declared no conflict of interest.

\bibliography{ivflr_ref}

\appendix

\bmsection{Table for Simulation Study 3\label{app1}}
This section presents the detailed results of Simulation Study 3.

\begin{center}
\setlength{\footnotesize\tabcolsep}{3.5pt} % default value: 6pt
\begin{longtable}{rrrr|ccccc}
\caption{The influence of the variance-covariance structures (Struct.) and their correlations ($\rho$) for the functional error terms on the performance of five estimators (Oracle, MULTI-2SLS, PW-2SLS, SIMEX, and Naive) in estimating $\beta_{1}(t)$, $n = 1000$, the functional variables with $\varepsilon_{X}(t) \sim MVN(0, \Sigma_{X})$ with $\Sigma_{X} = 1.5$, $U(t) \sim MVN(0, \Sigma_{U})$ with $\Sigma_{U} = 1$, $\eta(t) \sim MVN(0, \Sigma_{M})$ with $\Sigma_{M} = 1$, $\delta(t) = 0.5 \sin(2 \pi t)+1$.}
\label{table:VCOV} \\
\hline
&&&&\multicolumn{5}{c}{$\text{ABias}^{2}$ (MSPEE, \%)}\\
\hline
Struct.&$\rho_{X}$&$\rho_{W}$&$\rho_{M}$&Oracle&MULTI-2SLS&PW-2SLS&SIMEX&Naive\\
\hline
IND & 0.00 & 0.00 & 0.00 & 0.0013 (5.0) & 0.0013 ( 4.9) & 0.0014 ( 5.0) & 0.0057 (10.6) & 0.0485 (31.3)\\
AR1 & 0.50 & 0.50 & 0.50 & 0.0005 (2.9) & 0.0003 ( 2.4) & 0.0004 ( 2.7) & 0.0047 ( 9.7) & 0.0468 (30.7)\\
AR1 & 0.25 & 0.50 & 0.50 & 0.0019 (6.0) & 0.0017 ( 5.6) & 0.0290 (24.1) & 0.0224 (21.2) & 0.0984 (44.6)\\
AR1 & 0.75 & 0.50 & 0.50 & 0.0001 (1.2) & 0.0001 ( 1.2) & 0.0535 (32.9) & 0.0003 ( 2.4) & 0.0136 (16.6)\\
AR1 & 0.50 & 0.25 & 0.50 & 0.0001 (1.5) & 0.0001 ( 1.4) & 0.0001 ( 1.5) & 0.0008 ( 4.1) & 0.0198 (20.0)\\
AR1 & 0.50 & 0.75 & 0.50 & 0.0025 (6.8) & 0.0024 ( 6.6) & 0.0024 ( 6.7) & 0.0370 (27.3) & 0.1248 (50.2)\\
AR1 & 0.50 & 0.50 & 0.25 & 0.0005 (2.9) & 0.0004 ( 2.7) & 0.0276 (23.5) & 0.0045 ( 9.5) & 0.0468 (30.7)\\
AR1 & 0.50 & 0.50 & 0.75 & 0.0005 (2.9) & 0.0004 ( 2.6) & 0.0510 (32.1) & 0.0046 ( 9.6) & 0.0468 (30.7)\\
CS & 0.50 & 0.50 & 0.50 & 0.0024 (6.7) & 0.0024 ( 6.7) & 0.2834 (75.5) & 0.0034 ( 8.2) & 0.0491 (31.5)\\
CS & 0.25 & 0.50 & 0.50 & 0.0016 (5.4) & 0.0015 ( 5.2) & 0.1690 (58.3) & 0.0011 ( 4.6) & 0.0275 (23.5)\\
CS & 0.75 & 0.50 & 0.50 & 0.0035 (8.0) & 0.0038 ( 8.5) & 0.3847 (88.0) & 0.0291 (24.2) & 0.1130 (47.8)\\
CS & 0.50 & 0.25 & 0.50 & 0.0028 (7.3) & 0.0029 ( 7.3) & 0.2830 (75.4) & 0.0182 (19.1) & 0.0815 (40.6)\\
CS & 0.50 & 0.75 & 0.50 & 0.0018 (5.7) & 0.0018 ( 5.9) & 0.2806 (75.1) & 0.0036 ( 8.4) & 0.0179 (19.0)\\
CS & 0.50 & 0.50 & 0.25 & 0.0024 (6.7) & 0.0023 ( 6.6) & 0.2218 (66.8) & 0.0037 ( 8.5) & 0.0491 (31.5)\\
CS & 0.50 & 0.50 & 0.75 & 0.0024 (6.7) & 0.0024 ( 6.7) & 0.3167 (79.8) & 0.0030 ( 7.6) & 0.0491 (31.5)\\
UN & 0.50 & 0.50 & 0.50 & 0.0042 (8.9) & 0.0053 (10.1) & 0.2096 (64.6) & 0.0210 (20.5) & 0.0656 (36.4)\\
UN & 0.25 & 0.50 & 0.50 & 0.0048 (9.5) & 0.0058 (10.6) & 0.1326 (51.3) & 0.0159 (17.8) & 0.0530 (32.7)\\
UN & 0.75 & 0.50 & 0.50 & 0.0034 (8.0) & 0.0040 ( 8.7) & 0.2673 (73.0) & 0.0319 (25.3) & 0.0819 (40.6)\\
UN & 0.50 & 0.25 & 0.50 & 0.0042 (8.9) & 0.0048 ( 9.6) & 0.2087 (64.5) & 0.0237 (21.8) & 0.0769 (39.4)\\
UN & 0.50 & 0.75 & 0.50 & 0.0046 (9.3) & 0.0057 (10.5) & 0.2093 (64.5) & 0.0151 (17.4) & 0.0556 (33.5)\\
UN & 0.50 & 0.50 & 0.25 & 0.0042 (8.9) & 0.0052 (10.0) & 0.1507 (54.8) & 0.0210 (20.5) & 0.0656 (36.4)\\
UN & 0.50 & 0.50 & 0.75 & 0.0042 (8.9) & 0.0051 ( 9.9) & 0.2433 (69.6) & 0.0201 (20.0) & 0.0656 (36.4)\\
\hline
&&&&\multicolumn{5}{c}{AVar}\\
\hline
Struct.&$\rho_{X}$&$\rho_{W}$&$\rho_{M}$&Oracle&MULTI-2SLS&PW-2SLS&SIMEX&Naive\\
\hline
IND & 0.00 & 0.00 & 0.00 & 0.0040 & 0.0075 & 0.0072 & 0.0061 & 0.0032\\
AR1 & 0.50 & 0.50 & 0.50 & 0.0020 & 0.0050 & 0.0043 & 0.0039 & 0.0019\\
AR1 & 0.25 & 0.50 & 0.50 & 0.0028 & 0.0092 & 0.0044 & 0.0047 & 0.0022\\
AR1 & 0.75 & 0.50 & 0.50 & 0.0007 & 0.0018 & 0.0031 & 0.0027 & 0.0011\\
AR1 & 0.50 & 0.25 & 0.50 & 0.0016 & 0.0035 & 0.0040 & 0.0043 & 0.0017\\
AR1 & 0.50 & 0.75 & 0.50 & 0.0020 & 0.0071 & 0.0043 & 0.0035 & 0.0017\\
AR1 & 0.50 & 0.50 & 0.25 & 0.0020 & 0.0043 & 0.0049 & 0.0034 & 0.0019\\
AR1 & 0.50 & 0.50 & 0.75 & 0.0020 & 0.0064 & 0.0036 & 0.0051 & 0.0019\\
CS & 0.50 & 0.50 & 0.50 & 0.0055 & 0.0197 & 0.0076 & 0.0356 & 0.0041\\
CS & 0.25 & 0.50 & 0.50 & 0.0042 & 0.0108 & 0.0070 & 0.0269 & 0.0035\\
CS & 0.75 & 0.50 & 0.50 & 0.0086 & 0.0694 & 0.0089 & 0.0345 & 0.0050\\
CS & 0.50 & 0.25 & 0.50 & 0.0050 & 0.0215 & 0.0073 & 0.0262 & 0.0036\\
CS & 0.50 & 0.75 & 0.50 & 0.0058 & 0.0174 & 0.0078 & 0.0517 & 0.0052\\
CS & 0.50 & 0.50 & 0.25 & 0.0055 & 0.0165 & 0.0068 & 0.0310 & 0.0041\\
CS & 0.50 & 0.50 & 0.75 & 0.0055 & 0.0238 & 0.0087 & 0.0379 & 0.0041\\
UN & 0.50 & 0.50 & 0.50 & 0.0032 & 0.0120 & 0.0089 & 0.0255 & 0.0037\\
UN & 0.25 & 0.50 & 0.50 & 0.0022 & 0.0081 & 0.0058 & 0.0182 & 0.0023\\
UN & 0.75 & 0.50 & 0.50 & 0.0043 & 0.0188 & 0.0132 & 0.0320 & 0.0055\\
UN & 0.50 & 0.25 & 0.50 & 0.0031 & 0.0135 & 0.0095 & 0.0227 & 0.0037\\
UN & 0.50 & 0.75 & 0.50 & 0.0029 & 0.0111 & 0.0082 & 0.0284 & 0.0033\\
UN & 0.50 & 0.50 & 0.25 & 0.0032 & 0.0102 & 0.0077 & 0.0223 & 0.0037\\
UN & 0.50 & 0.50 & 0.75 & 0.0032 & 0.0136 & 0.0111 & 0.0276 & 0.0037\\
\hline
&&&&\multicolumn{5}{c}{AIMSE}\\
\hline
Struct.&$\rho_{X}$&$\rho_{W}$&$\rho_{M}$&Oracle&MULTI-2SLS&PW-2SLS&SIMEX&Naive\\
\hline
IND & 0.00 & 0.00 & 0.00 & 0.0054 & 0.0088 & 0.0086 & 0.0118 & 0.0516\\
AR1 & 0.50 & 0.50 & 0.50 & 0.0025 & 0.0053 & 0.0047 & 0.0085 & 0.0487\\
AR1 & 0.25 & 0.50 & 0.50 & 0.0047 & 0.0109 & 0.0334 & 0.0271 & 0.1006\\
AR1 & 0.75 & 0.50 & 0.50 & 0.0008 & 0.0019 & 0.0567 & 0.0029 & 0.0147\\
AR1 & 0.50 & 0.25 & 0.50 & 0.0017 & 0.0036 & 0.0041 & 0.0051 & 0.0215\\
AR1 & 0.50 & 0.75 & 0.50 & 0.0045 & 0.0095 & 0.0067 & 0.0405 & 0.1266\\
AR1 & 0.50 & 0.50 & 0.25 & 0.0025 & 0.0047 & 0.0324 & 0.0079 & 0.0487\\
AR1 & 0.50 & 0.50 & 0.75 & 0.0025 & 0.0068 & 0.0547 & 0.0097 & 0.0487\\
CS & 0.50 & 0.50 & 0.50 & 0.0078 & 0.0220 & 0.2909 & 0.0390 & 0.0532\\
CS & 0.25 & 0.50 & 0.50 & 0.0058 & 0.0122 & 0.1760 & 0.0280 & 0.0310\\
CS & 0.75 & 0.50 & 0.50 & 0.0120 & 0.0732 & 0.3936 & 0.0636 & 0.1180\\
CS & 0.50 & 0.25 & 0.50 & 0.0079 & 0.0243 & 0.2903 & 0.0445 & 0.0851\\
CS & 0.50 & 0.75 & 0.50 & 0.0076 & 0.0192 & 0.2884 & 0.0552 & 0.0231\\
CS & 0.50 & 0.50 & 0.25 & 0.0078 & 0.0188 & 0.2287 & 0.0346 & 0.0532\\
CS & 0.50 & 0.50 & 0.75 & 0.0078 & 0.0262 & 0.3254 & 0.0409 & 0.0532\\
UN & 0.50 & 0.50 & 0.50 & 0.0074 & 0.0173 & 0.2185 & 0.0465 & 0.0692\\
UN & 0.25 & 0.50 & 0.50 & 0.0071 & 0.0140 & 0.1383 & 0.0341 & 0.0553\\
UN & 0.75 & 0.50 & 0.50 & 0.0077 & 0.0228 & 0.2806 & 0.0639 & 0.0874\\
UN & 0.50 & 0.25 & 0.50 & 0.0073 & 0.0183 & 0.2182 & 0.0464 & 0.0806\\
UN & 0.50 & 0.75 & 0.50 & 0.0075 & 0.0168 & 0.2174 & 0.0436 & 0.0588\\
UN & 0.50 & 0.50 & 0.25 & 0.0074 & 0.0154 & 0.1583 & 0.0433 & 0.0692\\
UN & 0.50 & 0.50 & 0.75 & 0.0074 & 0.0187 & 0.2544 & 0.0478 & 0.0692\\
\hline
\end{longtable}
\end{center}

\end{document}